\documentclass[structabstract]{aa}  
\usepackage{graphicx}
\usepackage{txfonts}
\begin{document}
   \title{Spectroscopic FIR mapping of the  disk and  galactic wind of M82  with Herschel--PACS.}

   \subtitle{}

   \author{A. Contursi
          \inst{1}
          \and
          A. Poglitsch\inst{1}
          \and
          J. Gr\'acia Carpio\inst{1}
          \and
          S. Veilleux \inst{2}
          \and
          E. Sturm\inst{1}
          \and
          J. Fischer\inst{3}
          \and
          A. Verma\inst{4}
          \and
          S. Hailey-Dunsheath \inst{1}
          \and
          D. Lutz \inst{1}
            \and
          R. Davies  \inst{1}
           \and
          E. Gonz\'alez-Alfonso  \inst{5}
           \and
          A. Sternberg \inst{6}
            \and
          R. Genzel \inst{1}
            \and
          L. Tacconi \inst{1}
          }

   \institute{Max-Planck-Institut f\"ur extraterrestrische Physik, Postfach 1312, 85741 Garching, Germany
              \email{contursi@mpe.mpg.de}
        \and
         Department of Astronomy, University of Maryland, College Park, MD 20742-2421
         \and
	 Naval Research Laboratory, Remote Sensing Division, 4555 Overlook Ave SW, Washington, DC 20375,USA
         \and
         Oxford University, Dept. of Astrophysics, Oxford OX1 3RH, UK
         \and
	 Universidad de Alcal\'a de Henares, Departamento de F\'{\i}sica, 
         Campus Universitario, 28871 Alcal\'a de Henares, Madrid, Spain         \and
         Tel Aviv University, Sackler School of Physics $\&$ Astronomy, Ramat Aviv 69978, Israel
 	 }

 
    \abstract
    {We present  maps of the main cooling lines of the
    neutral atomic gas ([OI] at 63 and 145 $\mu$m and
   [CII] at 158 $\mu$m) and in  the [OIII] 88 $\mu$m line of the starburst galaxy M82, 
   carried out with the PACS spectrometer on board the Herschel satellite.
 }
   { Our aim is to study  the nature of the neutral atomic
   gas of M82  and to compare this gas with the molecular and ionized gas in the M82 disk and outflow.}
   {By  applying   PDR modeling 
    we were able to derive maps of the main ISM physical parameters, including the
   optical depth ($\tau_{[CII]}$), at unprecedented spatial resolution ($\sim 300$ pc).} 
   {We can  clearly kinematically separate  the disk from  the outflow  in all lines.
    The $\tau_{[CII]}$  is less than 1 everywhere, is lower in the outflow than in the disk, and
    within the disk is lower in the starburst region.\\
   The [CII] and [OI] distributions are consistent with PDR emission both in the disk and in the outflow. 
    Surprisingly, in the outflow, the atomic and the ionized gas traced by the [OIII] line both have a deprojected 
    velocity of
    $\sim75~\rm{\rm{km~s^{-1}}}$, very similar to the average velocity of the outflowing  cold molecular gas
   ($\sim 100~\rm{\rm{km~s^{-1}}}$) and several times smaller than the outflowing material detected in $H\alpha$ ($\sim
   600~\rm{\rm{km~s^{-1}}}$). This suggests that the cold molecular and neutral atomic gas and  the  ionized 
   gas traced by the [OIII] 88 $\mu$m line are dynamically coupled to each
   other but decoupled from  the $H\alpha$ emitting gas. }
   {We propose a scenario where  cold  clouds from 
   the disk are entrained into  the outflow  by the winds where they  likely evaporate,  surviving  as  small, fairly dense cloudlets
   ($n_H\sim500-1000
   ~\rm{cm^{-3}}$, $G_0\sim500- 1000$, $T_{gas}\sim300~\rm{K}$).
   We show that the UV photons provided by the starburst are sufficient to excite  
     the  PDR shells around  the molecular cores  and probably also the ionized gas that flows at the same PDR velocity. 
   The mass of the neutral atomic
   component  in the outflow is    $\gtrsim 5-12\times  10^7~\rm{M_{\sun}}$  to be compared with that of the molecular component 
   ($3.3 \times 10^8~\rm{M_{\sun}}$) and  of the $H\alpha$ emitting gas ($5.8 \times 10^6~\rm{M_{\sun}}$). 
   The  mass loading factor, $\dot{\rm{M}}_{\rm{Outflow}}/\rm{SFR}$, of the molecular  plus neutral atomic gas in the outflow is  $\sim 2 $. 
   Energy and  momentum driven outflow models can explain the data equally well, 
   if all the  outflowing gas components are taken into account.
   }

   \maketitle
%

\section{Introduction}
In the last decades,  the primary role of galactic outflows 
on galaxy evolution has become evident (Heckman, Armus and Miley \cite{Heckman90}; McKeith
et al. \cite{McKeith95}, Lehnert and Heckman \cite{Lehnert95}, Heckman \cite{Heckman98}, Veilleux, Cecil and 
Bland-Hawthorn \cite{Veilleux05}, Springel,  Di Matteo and Hernquist \cite{Springel}). 
Outflows in galaxies powered by either  Active Galactic Nuclei (AGNs) or 
starbursts have been introduced in theoretical models (e.g. Croton et al. \cite{Croton}) 
as sources of negative feedback, that terminate black-hole growth and/or star-formation 
and to explain the tight correlation between the mass of the black hole,
and the velocity dispersion or mass of the  bulge in which it resides
 (Magorrian et al. \cite{Magorrian}, Gebhardt et al. \cite{Gebhardt}, Merritt and  Ferrarese
\cite{Merritt}, Tremaine et al. \cite{Tremaine}, G\"ultekin et al. \cite{Gultekin}). 
These outflows contain a hot (10$^7$ K)  metal-enriched wind, 
in addition to entrained cooler gas and dust. Therefore, they may significantly
influence the chemical  evolution of both    the interstellar
medium (ISM) of the  galaxy and the intergalactic medium (IGM)
where this processed material  might be eventually dispersed.
  
Nearly   all galaxies with high star
formation rates show signs of outflows (Rupke, Veilleux and Sanders 
\cite{Rupke05a}, \cite{Rupke05b}, \cite{Rupke05c}). 
They also seem to be ubiquitous in Lyman Break galaxies at $z$ $\sim$ 3 
where they are capable of creating  holes of $\sim$ 100 kpc in the surrounding IGM
(Pettini   et al. \cite{Pettini01}, Shapley et al. \cite{Shapley03}, Steidel et al. {\cite{Steidel10}, Genzel et al.
\cite{Genzel10}), in Lyman $\alpha$ emitting
galaxies at z $\sim$ 4 (Finkelstein et al. \cite{Finkelstein}) and   compact gas rich Damped Lyman-$\alpha$ at z= 2.2 (Noterdaeme
et al. {\cite{Noterdaeme}).
Recently,  very massive molecular and neutral gas AGN-driven outflows   have
been discovered in Mrk 231 (Fischer et al. \cite{Fischer10}, Feruglio et al. {\cite{Feruglio} , Rupke and
Veilleux \cite{Rupke11}, Cicone et al. \cite{Cicone}),  in the local early type galaxy NGC 1266  (Aalto et al. \cite{Aalto}),   
in a $z=2$ QSO (Wei\ss~ et al. \cite{Weiss2012}) and in others Ultra Luminous Infrared Galaxies (ULIRGs) (Sturm et al. \cite{Sturm11}  and
Chung et al. \cite{Chung}). Sturm et al. (\cite{Sturm11})  provided the first 
observational evidence that molecular outflows in some ULIRGs are powered by the AGN and that the mass outflow rates
of the outflowing material exceed the Star Formation Rate (SFR) of the host galaxies, confirming the theoretically postulated
fundamental role of the outflows on galaxy evolution.
 
The evident cosmological importance of this phenomenon has brought many
astronomers to study a few nearby galactic winds in
great detail. M82 is the best studied case.
It is the closest  (at a distance of 3.6 Mpc, Freedman et
al. \cite{Freedman}) and the brightest  galaxy  
hosting a spectacular  bipolar outflow along its minor axis, although not as massive
and energetic as those recently discovered in ULIRGs. Nevertheless, thanks to its proximity
and to  its favorable disk inclination   ($i$ $=$ 81$^{\circ}$), M82 is one
 of the nearby galaxies most studied at all wavelengths and has therefore become 
 the archetype of a starburst with an outflow, with no AGN.
 
M82 has undergone  two powerful starburst  episodes
located in the central $\sim$ 500 pc of the galaxy.
 The first starburst was  triggered  10$^7$ yr ago 
 by the encounter with M81, 10$^8$yr  ago. 
 The second starburst was most likely bar driven 5$\times$10$^6$ yr ago 
 (F\"orster Schreiber et al. \cite{Natascha03}). Both of these starbursts comprise  several star
 clusters each hosting many hundreds of young massive stars (Melo et al. \cite{Melo}).
The superwind which feeds the bipolar outflow along the minor axis of the galaxy
is generated  in the inner 350 pc of  the galaxy where intense diffuse 
hard X-ray emission is detected (Strickland and Heckman \cite{Strickland}).
The outflow   reveals
filamentary structures at many wavelengths: they are   well visible in any high
spatial  resolution observation such
as  
in the  optical   H$\alpha$, [NII] and [OIII] lines (Shopbell and Bland--Hawthorn  \cite{Shopbell}), 
Xray (Strickland and Heckman \cite{Strickland}), warm molecular hydrogen (Veilleux, Rupke and  Swaters \cite{H2}) and  PAHs 
(Engelbracht et al. \cite{PAH}).
Cold and warm dust and molecular  gas are found in a biconical-like structure extending up to 
3 kpc away from the stellar disk on both sides, as revealed by 
 UV scattered light  (Hoopes et al. {\cite{Hoopes}), PAH emission (Engelbracht et
al.  \cite{PAH}, Kaneda et al. \cite{Kaneda}) sub--millimeter continuum emission  (Leeuw and  Robson \cite{Leeuw}
, Roussel et al. \cite{Roussel}), $^{12}$CO ($J$ 1$\rightarrow$0)  (Walter et al. \cite{Walter}, 
Wei\ss, Walter and Scoville \cite{Weiss}), 
 $^{12}$CO ($J$
3$\rightarrow$2)  (Seaquist and Clark \cite{Seaquist}) line emission and 
 H$_2$ near infrared observations (Veilleux,  Rupke and  Swaters \cite{H2}).   Most likely, this material has    been 
  entrained by the wind. Further out, most of the 
  dust (Roussel et al. \cite{Roussel}) and gas (Yun et al. \cite{Yun}) associated with 
  the halo seems to originate  from the tidal interaction with M81. \\ 
The  bibolar outflow  shows  different morphologies in the north and in the south, especially at optical wavelengths
(Shopbell and Bland--Hawthorn, \cite{Shopbell}, Westmoquette et al.
\cite{Westmoquette09a}, \cite{Westmoquette09b}, 
Seaquist and Clark \cite{Seaquist}). The southern wind is approaching us and is more 
clearly visible because the inclination of M82's disk provides a more direct sight-line 
to the southern part of the disk. 
The northern outflow is receding from us and is likely to be more heavily obscured by
the disk itself. This is why most detailed studies of M82's outflow have focused on the southern  
rather than the northern outflow of M82
(Westmoquette et el. \cite{Westmoquette09a}, \cite{Westmoquette09b}, Shopbell and  Bland-Hawthorn
 \cite{Shopbell}).\\
The most accepted interpretation of the outflowing material  in M82 is the following: 
 the supernova  explosions and stellar winds of the starburst  blow  away a  collimated hot ($\sim 10^{8}$ K) gas,
detected in the center of the galaxy in X-ray (Griffiths et al. \cite{Griffiths}).
The wind also drags cold material from the disk into the 
outflow. The collision between the hot fluid and the entrained cold clouds produces X-ray emission further out along
the outflow. These clouds are   heated   by shocks and the UV
radiation from the starburst producing most of the   
$H\alpha$ emission. 
The cold material in the winds has been   observed
in its dust and molecular component and  is  directly  illuminated 
 by the light of the starburst for at least $\sim 1~ kpc$ in both directions  (Veilleux,  Rupke and  Swaters
 \cite{H2}).\\
So far little observations are available on the neutral atomic component of the  outflow.
Observations in the HI line of the M81/M82 group revealed many tidal  streams
from M82 (Yun et al. \cite{Yun}, Chynoweth et al. \cite{Chynoweth}), 
but no detailed study has been conducted on the HI associated with
the winds. Colbert et al. (\cite{Colbert}) have 
studied the far infrared
(FIR) fine structure lines observed with the $LWS$ instrument on board of the $Infrared~Space~Observatory $
($ISO$)  but the spatial   ($\sim 80\arcsec$) and spectral ($\sim 10^3~ {\rm km~s^{-1}}$}) resolution were too poor to
distinguish the outflow and the disk components.
This is why we observed M82 with the PACS spectrometer 
(Poglitsch et al. \cite{Albrecht}) on board of the Herschel satellite (Pilbratt et al. \cite{Pilbratt})
in the [OI] lines at 63.2 and 145.6 $\mu$m, and the [NII], [CII] and [OIII] lines at 121.89, 157.7 and 88.3 $\mu$m 
respectively,
as part of the $SHINING$ Guaranteed Time Key Project  (P.I. E. Sturm). 
With these observations we were able to resolve for the first time {\it both
 spatially and kinematically} the disk and the outflow of M82 in the main FIR fine structure lines.\\
This paper is organized as follows: details on the observations and data reduction are given in Section 2.
The resulting maps and related uncertainties are presented in Section 3. In Section 4 
we explain the Photo-dissociated Regions (PDR) modeling of our observations and we discuss the meaning 
of the physical parameters 
resulting from this modeling. We also compare the various gas phases participating to the outflow  
and we discuss their  energetics and their relation.  
In Section 5 we summarize the main conclusions.

  \begin{figure*}
   \centering
  
\includegraphics[angle=0,width=15.5cm,height=19.0cm]{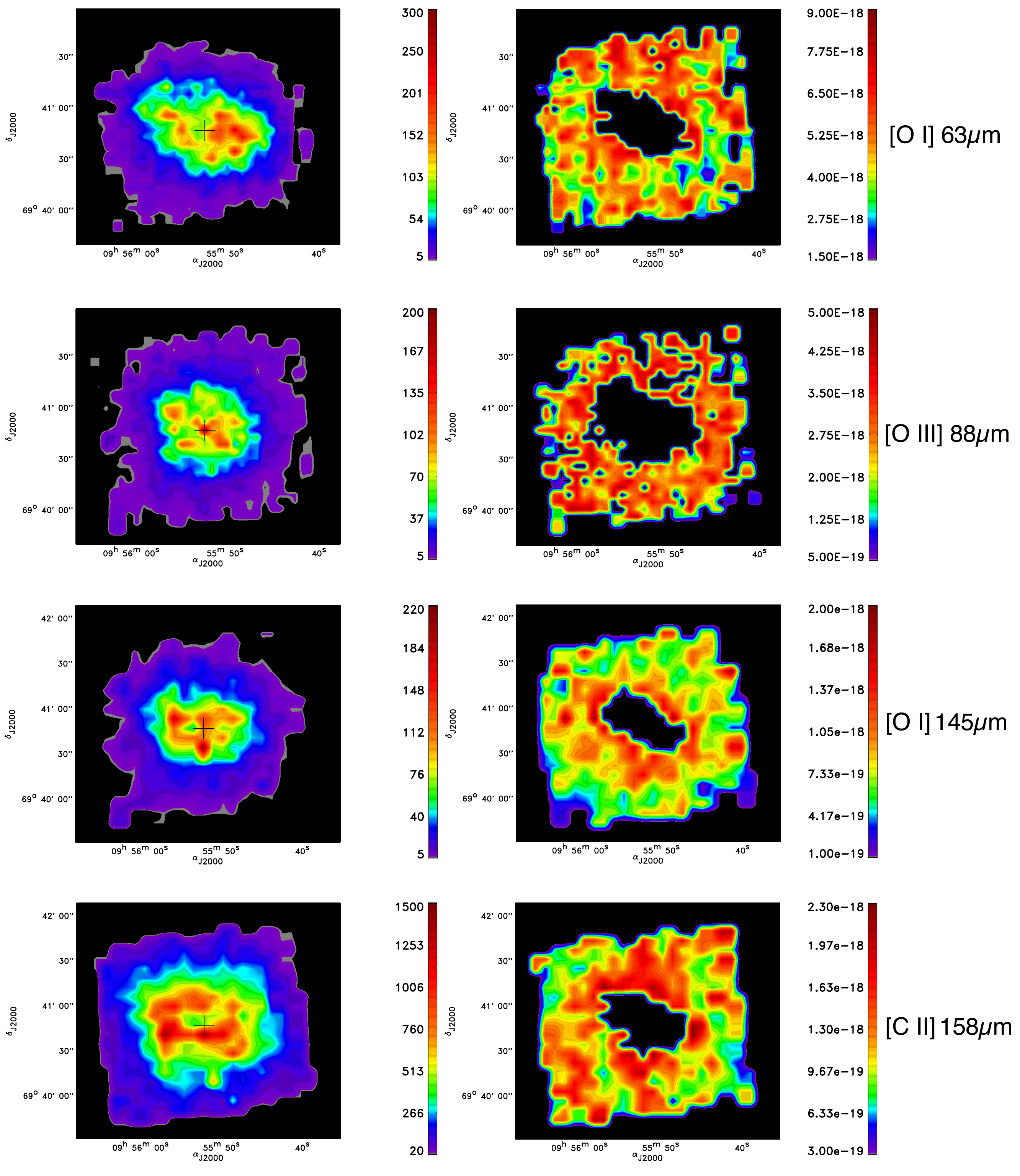}
       \caption{Left column: for each line the signal to noise ratios maps. Right column: for each line the   "error"
      maps in ${\rm W~m^{-2}}$, obtained as explained in Section 2 with the central region masked.
              }
         \label{errormaps}
   \end{figure*}


\section{PACS spectrometer observations and data reduction}
We have observed M82 with a 5$\times$5 (4$\times$4) raster in the blue (red)
channel, and with step sizes in YZ spacecraft coordinates equal to: $step_Z$ = 16.0$\arcsec$ (24.0$\arcsec$)  and 
$step_Y$ 14.5$\arcsec$ (22.0 $\arcsec$), in bright
line scan chopped mode, with large chopper throw in the following fine structure lines:
 [OI]
($^3$P$_2$$\rightarrow$$^3$P$_1$) and ($^3$P$_1$$\rightarrow$$^3$P$_0$) lines at
63.2 and 145.6 $\mu$m, [CII] ($^2$P$_{1/2}$$\rightarrow$$^3$P$_{3/2}$) line
at 157.7 $\mu$m and  [OIII] ($^3$P$_0$$\rightarrow$$^3$P$_1$)  line at 88.3
$\mu$m. 
 The on source integration times per raster position  are: $\sim 40, ~  40,~ 65, $ and $  33~$s 
for the  [OI] 63, [OIII] 88, [OI] 145 and [CII] 158 lines respectively. 
Althought the bright line mode executes less grating steps 
{\it per} scan  than the {\it faint mode},  it nevertheless attains Nyquist sampling. 
 The spectral resolutions  for the  
lines are: $\sim 3300$ (87 km s$^{-1}$), $\sim 2400$ (125 km s$^{-1}$), 1160 (257km s$^{-1}$) and  $\sim 1255$ (240 km
s$^{-1}$), respectively. \\

The spatial steps used in the rasters, in combination with the spaxel pattern of the instrument, ensure spatially  Nyquist sampled final maps.\\
For each line, each raster position  has been reduced in HIPE{\footnote{HIPE is a joint development by the Herschel Science Ground
Segment Consortium, consisting of ESA, the NASA Herschel Science
Center, and the HIFI, PACS and SPIRE consortia.}} with the standard pipeline up
to the rebinnedCube task. Then, a polynomial  continuum plus a Gaussian profile in each pixel at each raster position
have been fitted, and the corresponding intensity, wavelength peak, Full Width Half Maximum (FWHM) and 
celestial coordinates  were  recorded. In an external package
we built the final maps applying the drizzle algorithm (Fruchter and Hook \cite{Drizzle}).
Thus, we produced   intensity, peak velocity and line width  maps for the four observed
lines. Moreover, thanks to the brightness of M82 we were also able to produce 
high spatial resolution continuum maps using the continuum measured from our line observations 
as well as from the data in the parallel channels.
The final maps cover  the disk and the base 
of the outflow of M82 up to 1 kpc away from the disk in both direction, at an unprecendent spatial resolution (from 130
to 270 pc) and sensitivity  (see Figure 12 of Poglitsch et al. \cite{Albrecht}) at these wavelengths.
 Taking into account pointing offset and jittering during observations and the uncertainties related to the fact that we 
arbitrarily assign  the coordinates at the center rather than at the corner of the final pixel,
 we estimate an overall  astrometric uncertainty  of
$\sim$ 4 $\arcsec$.\\
In addition to the rasters, we have observed M82 also in the  ($^3$P$_1$$\rightarrow$$^3$P$_2$) 
[NII] line at 121.89 $\mu$m in faint chop-nod line scan mode, with 3 separate  staring observations pointed 
on the  galaxy's center, the northern and the southern outflow. 
These data  were reduced  in HIPE up to the rebinned cube step resulting in 25 line intensity values per pointing.
 
   \begin{figure*}
   \centering
  
\includegraphics[angle=0,width=17.5cm,height=13.0cm]{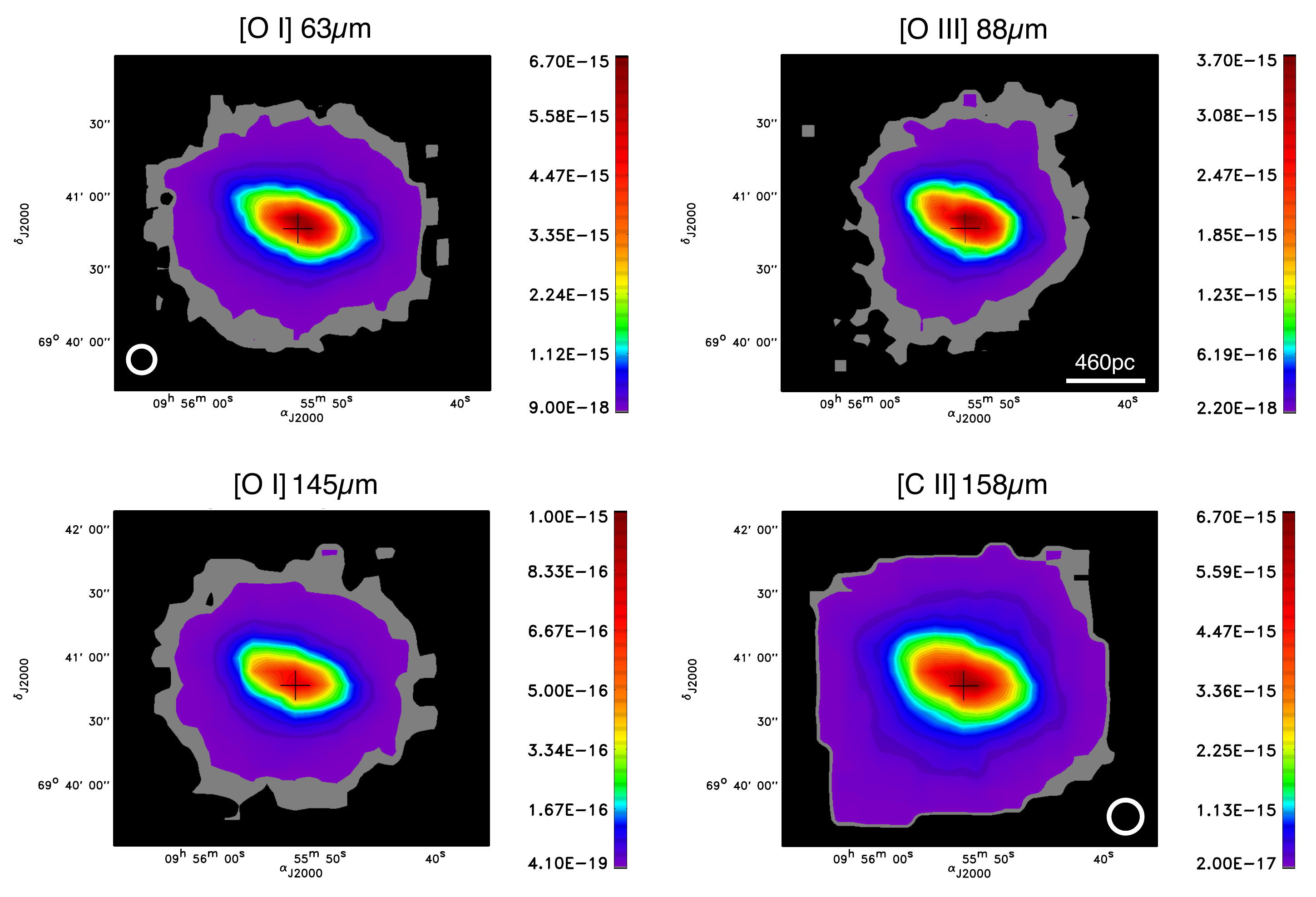}
      \caption{Integrated line emission of the 4 observed lines at their original  spatial  resolution.
      Typical  PSF widths are shown for the shortest  and longest wavelengths.  Units are in
      $\rm{W~m^{-2}}$. The black cross corresponds to the adopted center of M82,
      {\it i.e.}  $9h55m52.2s~ 69d40m46.6s$  (J2000) (Dietz et al
      \cite{DistM82}). North is up and East is towards the left.
              }
         \label{line_int}
   \end{figure*}
  
For each of the fitted spectra we have estimated the errors from the residual spectrum, $i.e.$ the original spectrum minus the fitted continuum
plus Gaussian profile. These errors contain the noise of the spectrum as well as systematic deviations from  a perfect fit.  These
errors were quoted in the same units as the fitted line fluxes, $i.e.$, $\rm{W~ m^{-2}}$ per each spaxel. The
line flux error maps were generated with exactly the same drizzle scheme as for the line map. 
In order  to create the error maps by the same method as for the flux, we used the intermediate quantity 1/errors$^2$  which
can be drizzled as an extensive variable.  
However, caution must be taken when interpreting the resulting  errors for the following reasons.
In principle, if we were dealing with statistical noise
only, by observing the same spot in
the sky several times, as we effectively do when we combine different
rasters in the drizzle, the error of the combined measurements would go
down. If there is a systematic error it will always be the same in sign
and magnitude and will not cancel with repeated measurements. Furthermore, the errors in the regions
 where the error/signal ratio is   better than  the  
calibration  accuracy (explained in detailed in Section 3.4)  should not be quoted as
absolute measurement errors. This explains why we have an increase of the estimated "errors"    where the signal is stronger.
Therefore, these maps are meaningful only in the "outskirts" of the
galaxy but not in the starburst region and for this reason   we masked the starburst regions in the "error" maps 
shown in the right column of Figure \ref{errormaps}.
In order to give a complete picture of the final "noise" levels we reach in each map, we show 
in  Figure  \ref{errormaps} also the S/N maps. 
}

\section{Results}
\subsection{FIR line intensity and continuum maps}
Figure \ref{line_int} shows the line intensity maps  of the 4 lines in their original spatial resolution. All maps
are North-South oriented.
The crosses represent the center of M82.
The brightest integrated line emissions are very similar at all wavelengths but
at the faint emission level, some important differences are noticeable.
The two lines tracing exclusively the neutral atomic gas, namely [OI] 63 $\mu$m and [OI 145] $\mu$m,    
show a  spherical morphology, while the [CII] emission, that  arises
mainly in the neutral  atomic medium but also from the ionized medium, shows a 
weak elongation  towards the minor axis. This elongation becomes more evident  in the [OIII] 88 $\mu$m line 
intensity maps, that traces purely ionized gas. In all maps, emission up to 1 kpc on both sides 
of the disk along the minor axis  is detected.
From these maps it is evident that the ionized medium has an emission elongated along the minor axis of
the galaxy while the emission arising from the neutral  medium is  more   spherically  distributed, although still detected along the minor axis of the
galaxy.\\
Figure \ref{88_cont}  shows the continuum map derived from the observation of the [OIII] 88 $\mu$m 
emission line. The continuum maps at the other
wavelengths show a similar morphology and therefore  they are not shown here.
The continuum morphology is spherical and, at the faint level of the emission, 
does not show a marked asymmetry along the galaxy's minor axis. 
The   brightest continuum emission region  is  displaced $\sim$ 6$\arcsec$ West with respect to the center of the galaxy 
(black cross on the Figure). This is marginally larger than the final astrometric uncertainties of our maps ($\sim
4\arcsec$). The peak emission moves towards the east  with increasing wavelengths indicating that the western part is hotter than
the eastern part in the starburst region of M82.  

%
   \begin{figure}
   \centering
  
\includegraphics[angle=0,width=8.5cm,height=6cm]{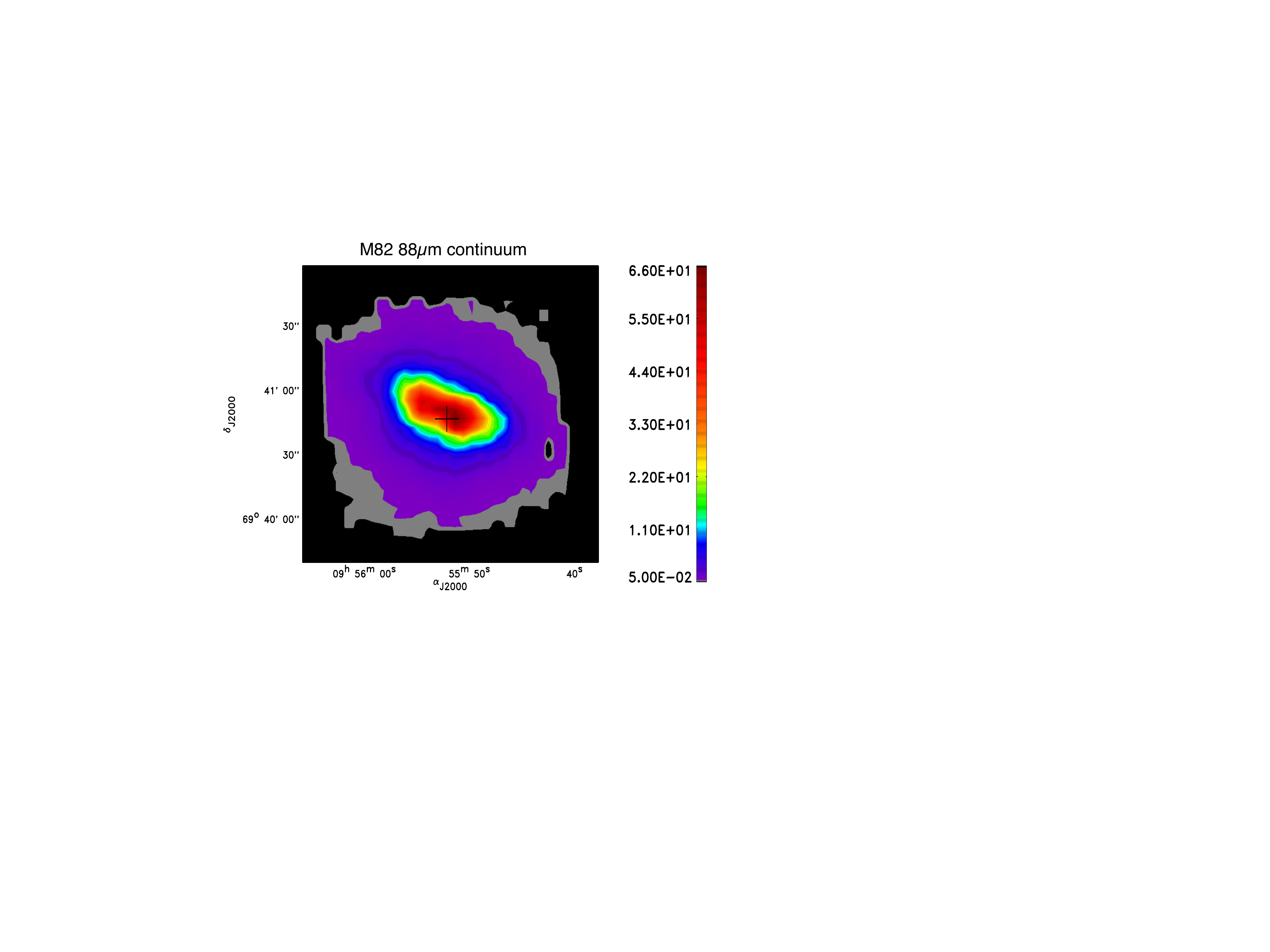}
      \caption{Offline continuum emission obtained from the continuum in  the [OIII] 88 $\mu$m line.  Units are in
     Jy. The black cross corresponds to the adopted center of M82.
              }
         \label{88_cont}
   \end{figure}

\subsection{Velocity dispersion maps}
Figure \ref{fwhm_int} shows  the  velocity dispersion maps obtained from  FWHM  of the line fits,     
converted in velocity and corrected for the instrumental  profile.
The overall shape is similar in all lines and it is formed by two well 
distinct kinematic  structures: the disk  and the outflow.
In general the two cones of the outflow have a velocity dispersion  higher (from $\sim 50$ to $\sim 100 
~\rm{km~s^{-1}}$ or a factor from 1.5 to 2) than  the disk.\\
All four maps clearly show  emission in the outflow aligned along the minor axis. 
This enhanced emission originates from a point that coincides with the center of the disk 
(as marked by the black cross in the figure).
The  yellow ellipse shown in the Figure
corresponds to the stellar disk defined by the emission in the $K_s$-band, as shown in
Figure \ref{line_ratio} and explained in Section 3.5.
 
   \begin{figure*}
   \centering
  
\includegraphics[angle=0,width=17.5cm,height=13.0cm]{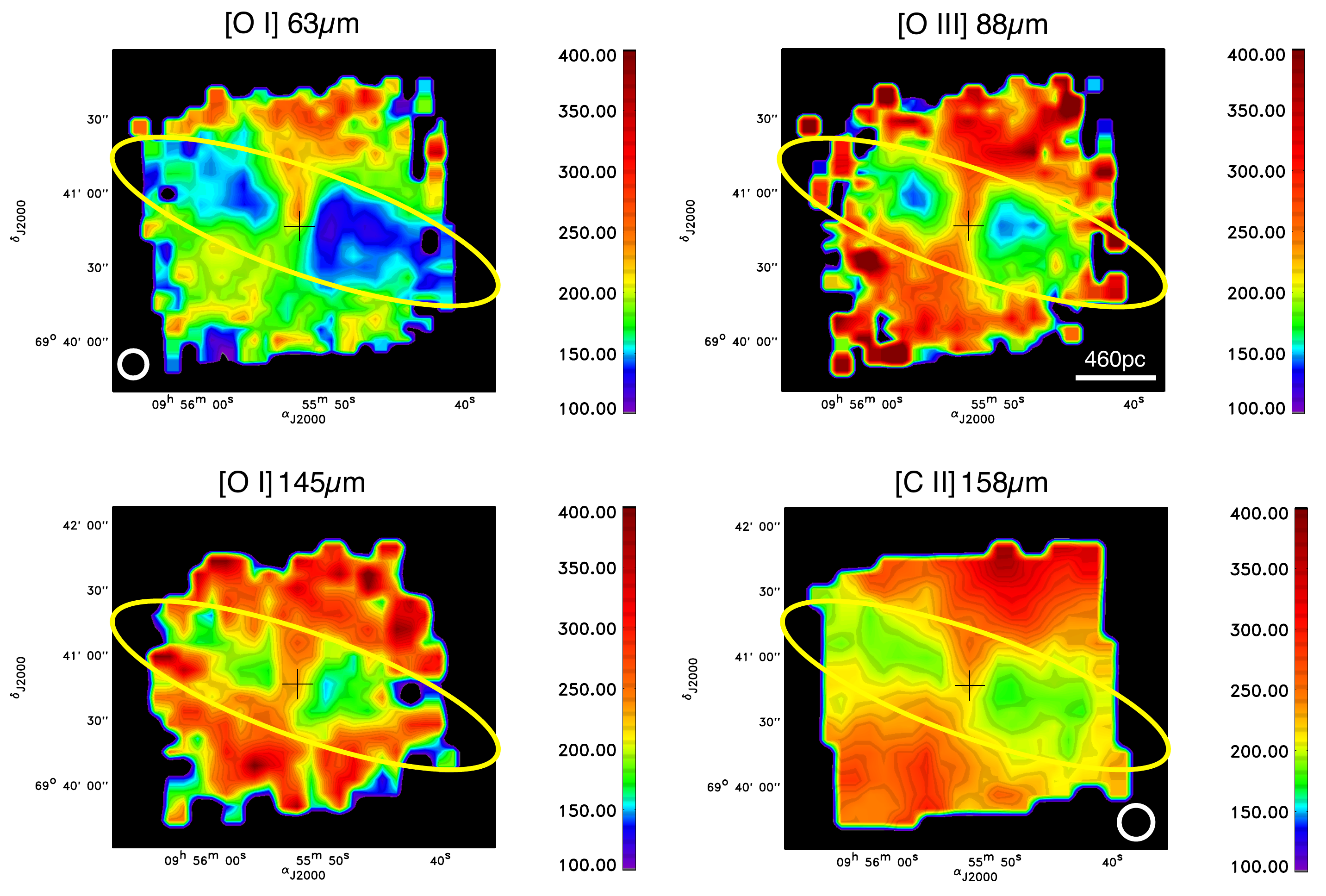}
      \caption{Velocity dispersion maps obtained from the fitted line widths deconvolved with the
      instrumental profile width, in $\rm{km~s^{-1}}$. The black cross corresponds to the adopted center of M82. The yellow
      ellipse corresponds to the stellar disk as described by the $K_s$-band  emission shown in contours  in Figure
      \ref{line_ratio}.}
         \label{fwhm_int}
   \end{figure*}

\subsection{Radial velocity maps}
Figure \ref{radial_int}   shows  the radial  velocity maps, obtained from the wavelength  corresponding to 
the fitted lines peak   
converted into velocity, from which we have subtracted the systemic velocity of M82 equal
 to  203 $\rm{km~s^{-1}}$ (Shopbell and Bland--Hawthorn \cite{Shopbell}).
The overall S--shape morphology of the velocity field is similar in all lines and confirms previous
findings:  the  east part of the disk and the north outflow are receding from us while the west part of the disk 
and the southern outflow are approaching us. The region where the velocity is zero is consistent in
all 4 maps and it corresponds to the galaxy's center (black crosses).
The velocity in the disk goes from  $\sim$-80 to $100~\rm{km~s^{-1}}$ in all lines but the [OI] 145 $\mu$m line,
which is shifted by $\sim$+20 $\rm{km~s^{-1}}$ with respect to the others. 
This shift is present in all raster positions suggesting a systematic 
error (see next section for  details) rather than a real physical effect.
We have also derived  rotation curves along the major axis of the
galaxy (see Figure \ref{rotcurves}). All curves are consistent with each other (except for the velocity shift in 
the rotation curve obtained with the  [OI] 145 $\mu$m line) and they  also agree with those shown
 by Shopbell and Bland--Hawthorn (\cite{Shopbell}), especially with those  in H$\alpha$ and $^ {12}$$CO(J 2\rightarrow 1)$.\\
  It is important to note that {\it we do not detect line splitting in any of the observed  lines}. 
 This is particularly intriguing  for the [OIII] line at 88 $\mu$m.
 Indeed, line splitting   is clearly detected in $H\alpha$ in the outflow, above 300 pc from the galaxy plane along the minor 
 axis of the galaxy, with velocity separation of $\sim 200~\rm{km~s^{-1}}$  (Greve et al. \cite{Greve}),
 well above    the PACS
 spectral resolution of the [OIII] line at 88 $\mu$m (120 $\rm{km~s^{-1}}$).  
 It is not clear whether line splitting is detected in the optical 
 [OIII] lines: Shopbell and Bland-Hawthorn
   (\cite{Shopbell})  report    marginally detected [O III] 5007 $\AA$ line splitting   
   in some regions of the mapped area   but the low S/N prevented them to fit double Gaussians to their [O III] 5007 $\AA$ data
   and therefore to be more quantiative on this issue.
 As  will be discussed later, this is not the only striking difference we find
 between  the gas emitting in the [OIII] line at 88 $\mu$m and that emitting in H$\alpha$.\\

From the [OIII]/[OI] 63 $\mu$m map we will present in Section 3.5, where the bipolar outflow  can be best
determined, we measure an opening angle of $\sim$ 50$^{\circ}$ for both sides of the outflow, very similar
to the value found by Walter, Wei\ss~ and Scoville (\cite{Walter}) in the CO maps.
Taking into account  the disk inclination of M82 ($ i = 81$ ), the outflow  opening  angle, and assuming a simple jet-like
geometry,
we can derive the mean de-projected velocity dividing the observed velocity ($\sim 40-45 ~\rm{km~s^{-1}}$) by $sin(81^{\circ}-50^{\circ}/2)$.
This results in a mean deprojected velocity  from $\sim$ 75 to 85  $\rm{km~s^{-1}}$, similar within the wavelength 
uncertainties, in both cones of the outflow and at both wavelengths, {\it i.e.} for both the ionized gas traced by the [OIII] emission
and the neutral gas traced by the two [OI] lines and the [CII] line.

   \begin{figure*}
   \centering
  
\includegraphics[angle=0,width=17.5cm,height=13.0cm]{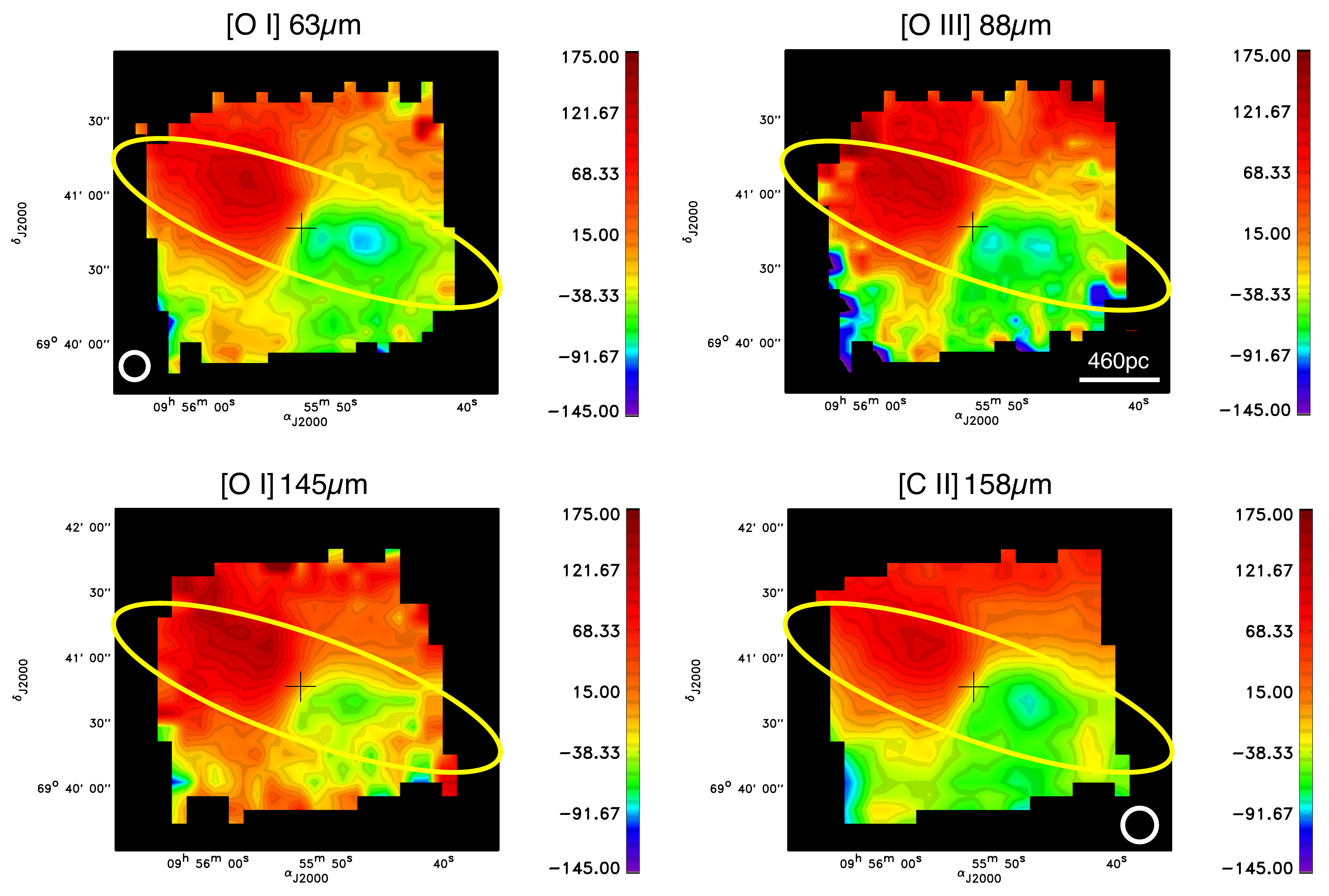}
     \caption{Radial velocity maps obtained from the fitted line profiles at   each
      position.  The black cross  corresponds to the adopted center of M82. }
         \label{radial_int}
   \end{figure*}
   
\subsection{Uncertainties}
 The dominant source of $flux~  uncertainties$ in our data are those related to the spectrometer flux calibration, that is
  $\sim$30$\%$ in both the blue and the red channels (Poglitsch   et
al. \cite{Albrecht}).
In the raster observations presented here, another source of error arises from  the fact that for all observed lines 
we used the {\it bright line} mode, {\it  i.e}  less grating steps per grating scan  than in faint line mode. 
The  reason for having chosen this observing mode    is that these maps were executed 
during the Performance Verification phase,  to check that the PSF sampling  scheme for the mappping mode worked as expected. 
In order to test it 
we had to observe as many  wavelengths as possible along the full PACS wavelength range ($55 - 220~ \rm{\mu m}$),
and faint line mode would have taken a prohibitive amount of time.\\
In bright line mode each of the 16 spectral pixels of one spatial pixel (hereafter spaxel) does not see the full line, 
as in the case of  faint line-scan. As a consequence, one of the crucial steps of the faint line mode data reduction, 
bringing the response of the 16 spectral pixels of each spaxel to the same continuum level (the so called FlatField
task) cannot be applied in the case of bright line mode.
Figure \ref{brightline} shows an example of a spectrum before rebinning and
combining the nods, for the central spaxel of
the central raster position at 63 $\mu$m. Each spectral pixel of the central spaxel has a different color. The figure
clearly shows how small is the overlapping spectral range seen by subsequent spectral pixels of each spaxel  and
illustrates the difficulties in flat fielding this type of observations.
The lack of the flat field could result in distorted line profiles, and therefore it could affect
the velocity peak, the FWHM and the total flux.\\
To  evaluate the impact of this particular observing mode on the intensity maps, we have  compared the LWS line flux
of M82 given in Brauher, Dale and Helou (\cite{Brauher}) and by Colbert et al. (\cite{Colbert}) with ours. We have assumed a circular Gaussian LWS beam profile, 
with a FWHM equal to 80\arcsec at all wavelengths but at 158 $\mu$m where we assumed a FWHM equal to
70\arcsec~ (Lloyd \cite{Lloyd}). We integrated the line  and
continuum PACS flux  under the assumed LWS profiles and we compared the obtained fluxes with those observed  with LWS  
 corrected for the LWS extended source 
correction factor (see Table 7 of Brauher, Dale and Helou \cite{Brauher}).
The results  are shown in Figure \ref{PACSLWS}. The LWS and PACS fluxes do agree within the
 respective uncertainties
adopted equal to  30$\%$ for PACS and 20$\%$ for LWS as claimed in Brauher, Dale
and Helou (\cite{Brauher}), with the possible exception of the continuum flux density at 88 $\mu$m which only marginally
agrees with the adopted uncertainties. We conclude that the bright line mode does not introduce uncertainties greater than
the nominal PACS spectrometer flux uncertainties.

The nominal uncertainties due to the $wavelength~ calibration$ are ~$\sim$ 20 $\rm{km~s^{-1}}$ and $\sim$ 40 $\rm{km~s^{-1}}$ 
in the blue and red channel respectively (Poglitsch  et al. \cite{Albrecht}). A distortion of the line
profile due to the bright line mode could introduce  further uncertainties which is difficult to quantify. 
However, we expect these
uncertainties to  increase with  decreasing   line intensity where the distortion of the lines profiles
become more significant. Among the four observed lines, the [OI] 145 $\mu$m
line is by far the weakest. Therefore, we expect the velocity line peak  of this line
 to be more affected than the others due to this observing mode.
 This could in principle explain the shift observed in the rotation curve of this line with respect to the rotation curves 
 obtained in the  other lines and visible in the mono dimensional rotation curves shown in
 Figure \ref{rotcurves}. On the other hand, the issues  with the bright-line mode would  modify the line profiles in a random way, 
 because the misalignment of the spectrum among the spectral pixels of the same spaxel depends on 
 the transient history of 
 each pixel and on the mispointing,  which has no systematic direction. Since the observed shift 
 in the [OI] 145 $\mu$m line is 
 systematic in all the raster positions, we believe that the problem arises from a wavelength calibration 
 issue at this wavelength.
We note however, that  the fact that the  rotation curves in the other lines agree with those 
previously published  indicates  that the bright line mode does not introduce significant
additional uncertainties  even at 145 $\mu$m   and that the data are dominated by the nominal wavelength calibration uncertainties.\\
The deformation of the line-shape due to the bright line mode is also a  random process and it would affect 
the fitted FWHM
from which we have derived the velocity dispersion maps shown in Figure \ref{fwhm_int}. The fact that all maps show the same
kinematic morphology, ensures that the line deformation does not affect the overall velocity dispersion in a systematic way,
producing fake structures.

   \begin{figure}
   \centering
  
\includegraphics[angle=0,width=8.5cm,height=7.5cm]{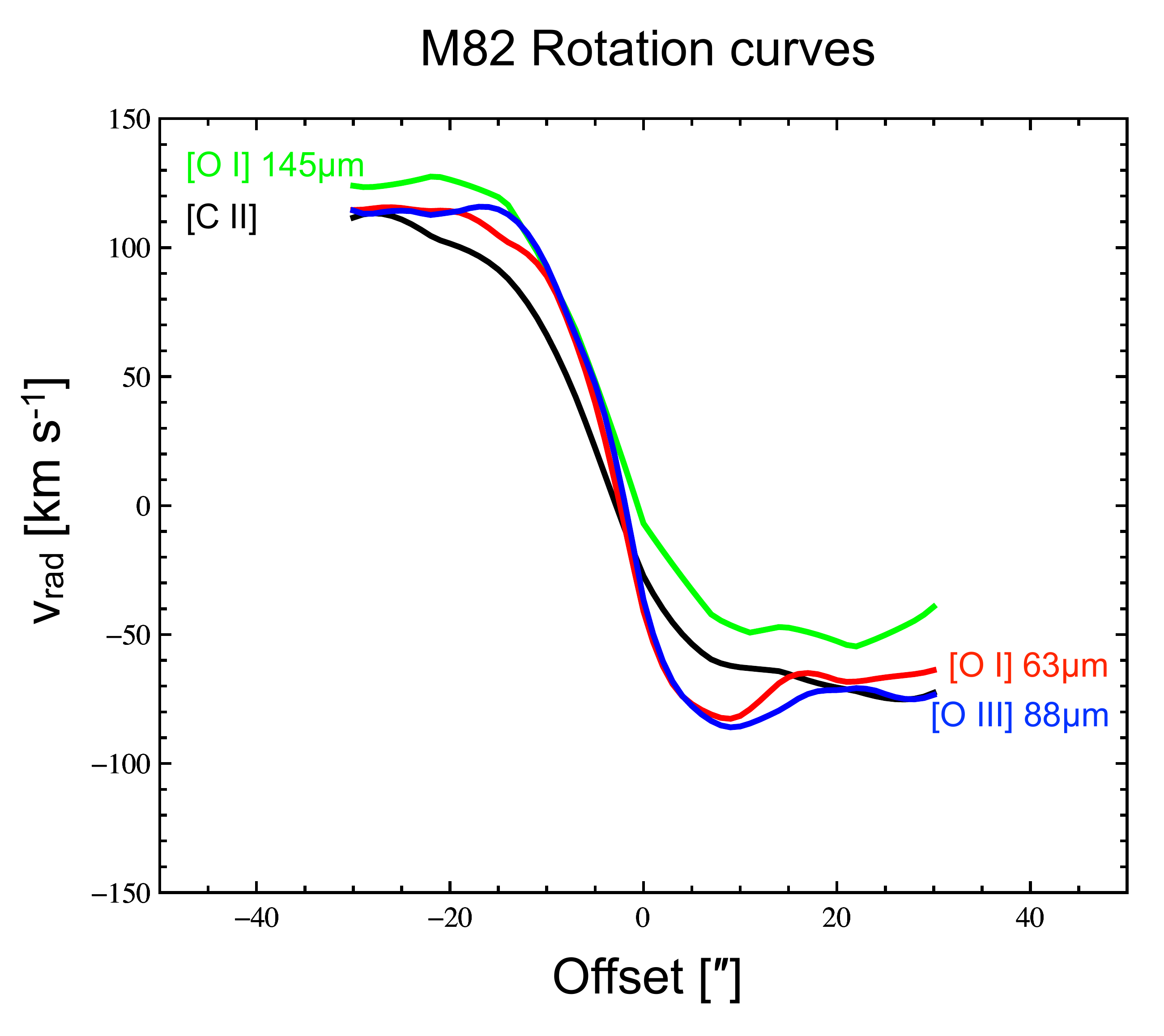}
      \caption{Rotation curves along the major axis of M82 in the 4 observed lines. Note the shift of the rotation
      curve obtained in the  [OI] 145$\mu$m line which is not a real physical effect (see section 3.4 for details). }
         \label{rotcurves}
   \end{figure}
 
   \begin{figure}
   \centering
  
\includegraphics[angle=0,width=8.0cm,height=8.0cm]{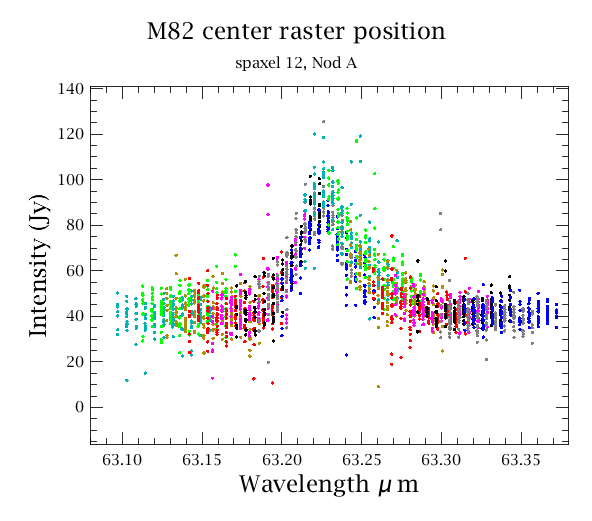}
      \caption{Example of bright line mode observation. The Figure shows the spectrum of all spectral pixel of the
      central spaxel of the field of view, in different color for each spectral pixel, for Nod A. }
         \label{brightline}
   \end{figure}

\subsection{Ratio Maps}
Figure \ref{line_ratio} shows the [OI] 63 $\mu$m /[CII], [OIII] /[OI] 63 $\mu$m, [OIII] /[CII] and [OI] 145$\mu$m /[OI] 63 $\mu$m ratio
maps. Before making each  ratio map, the two images have been smoothed  to the same resolution, if necessary.
The HST H$\alpha$ contours 
(Mutchler et al. {\cite{HST}) are overplotted on the [OIII] /[OI] 63 $\mu$m ratio map (see discussion  in the next Section).\\
The ratio maps are very different from each other: the [OI] 63 $\mu$m/[CII] shows an overall asymmetry along the disk major
axis, while in the [OIII]/[OI] 63 $\mu$m map and, at a poorer resolution also in the [OIII]/[CII]
ratio map,  the bipolar outflow is very clearly delineated. 
On the [OI] 63 $\mu$m/[CII]  ratio map in Figure \ref{line_ratio},  the K$_s$-band  contours (Veilleux, Rupke and Swaters
\cite{H2}) describing the stellar disk
are also plotted. Based on these contours we have defined two ellipses: one external (yellow  in Figure \ref{line_ratio})
describing the overall stellar disk
 and  one internal (purple in   Figure \ref{line_ratio}) that defines the brightest part
 of the stellar disk  including   the starburst region.
 
While the region where the [OI] 63 $\mu$m /[CII] ratio is high  corresponds well to this internal ellipse, the
region where the  [OIII]/[OI] 63 $\mu$m ratio is higher  arises from the  north  part of the starburst region. 
A comparison with the $^{12}$CO($J$ 1$\rightarrow$0) map (Figure 1 of Wei\ss, Walter and
Scoville  \cite{Weiss}) with the  [OIII]/[OI] 63 $\mu$m  ratio image shows that the outflow  coincides 
with the regions marked as outflow  in their figure.
The northern outflow is a factor of $\sim$ 2 brighter in the [OIII]/[OI] 63 $\mu$m ratio than the southern one. 
In contrast, at optical wavelengths the northern outflow has not been detected in the [OIII] emission line at 5007 $\AA$
   while the southern outflow is clearly visible at this wavelength (Figure 2 of Shopbell and Bland-Hawthorn
   \cite{Shopbell}). 
It seems unlikely that the difference in the observed [OIII] 5007 $\AA$ line emission in the two cones of the outflow of M82
is due to an electron  density  higher than the  critical density  of the  
[OIII] line at 5007 $\AA$  (6.3$\times 10^5$$\rm{cm^{-3}}$) in the 
northern outflow. In fact,   the estimated averaged electron density in the outflow  is
$\lesssim 300~ \rm{cm^{-3}}$ (Yoshida, Kawabata and Ohyama \cite{Yoshida}), and the average electron density associated with the 
filaments emitting in H$\alpha$   is $\sim 15~\rm{cm^{-3}}$  (Shopbell and Bland-Hawthorn
   \cite{Shopbell}). As we will argue in the next section, we believe
that this difference can be explained with the extinction being higher in the northern than in the southern outflow.

The  [OI] 145/[OI] 63 $\mu$m ratio is  shown in Figure \ref{line_ratio}. 
This ratio  varies only by a factor of 2  all over the
mapped area, and neither the  outflow  nor the stellar disk are clearly visible. 
In the starburst region defined by the purple ellipse,
this ratio appears slightly weaker than elsewhere. 
Figure 2 of Tielens and Hollenbach (\cite{Tielens85}) shows  how this ratio 
depends on the gas temperature for various gas densities. For gas temperatures higher than $\sim$300 K and 
a fixed gas density 
 this ratio is  basically constant and it decreases with increasing  density. 
The comparison with the range of ratios observed in  M82  ( $\sim 0.05-0.1$ ) suggests that the gas 
temperature is greater than $\sim$ 300 K everywhere and that
 the density is higher in the central region where this ratio is lower. We will see in Section 4.2 that more 
 detailed modeling confirms that this is the case.

Finally, Figure \ref{line_over_continuum} shows the line over the continuum as measured around the   [OIII] and
[OI]  63 $\mu$m lines. These figures confirm that both the neutral and the ionized gas traced by the FIR fine structure line
  participate in the outflow, but
that the ionized gas is more collimated than the neutral gas. This fits well in   the picture where the cold entrained
gas forms an outflow less collimated than the hot gas (Ohyama et al. \cite{Ohyama}). 
However, the ionized gas traced by the [OIII]
line at 88 $\mu$m is kinematically coupled with the neutral atomic gas and it is not kinematically coupled
 with the ionized gas traced by the $H\alpha$ emission (Greve \cite{Greve})
because it moves at a much lower velocity than the hot ($10^4$ K) gas.

   \begin{figure*}
   \centering
  
\includegraphics[angle=0,width=15cm,height=7.0cm]{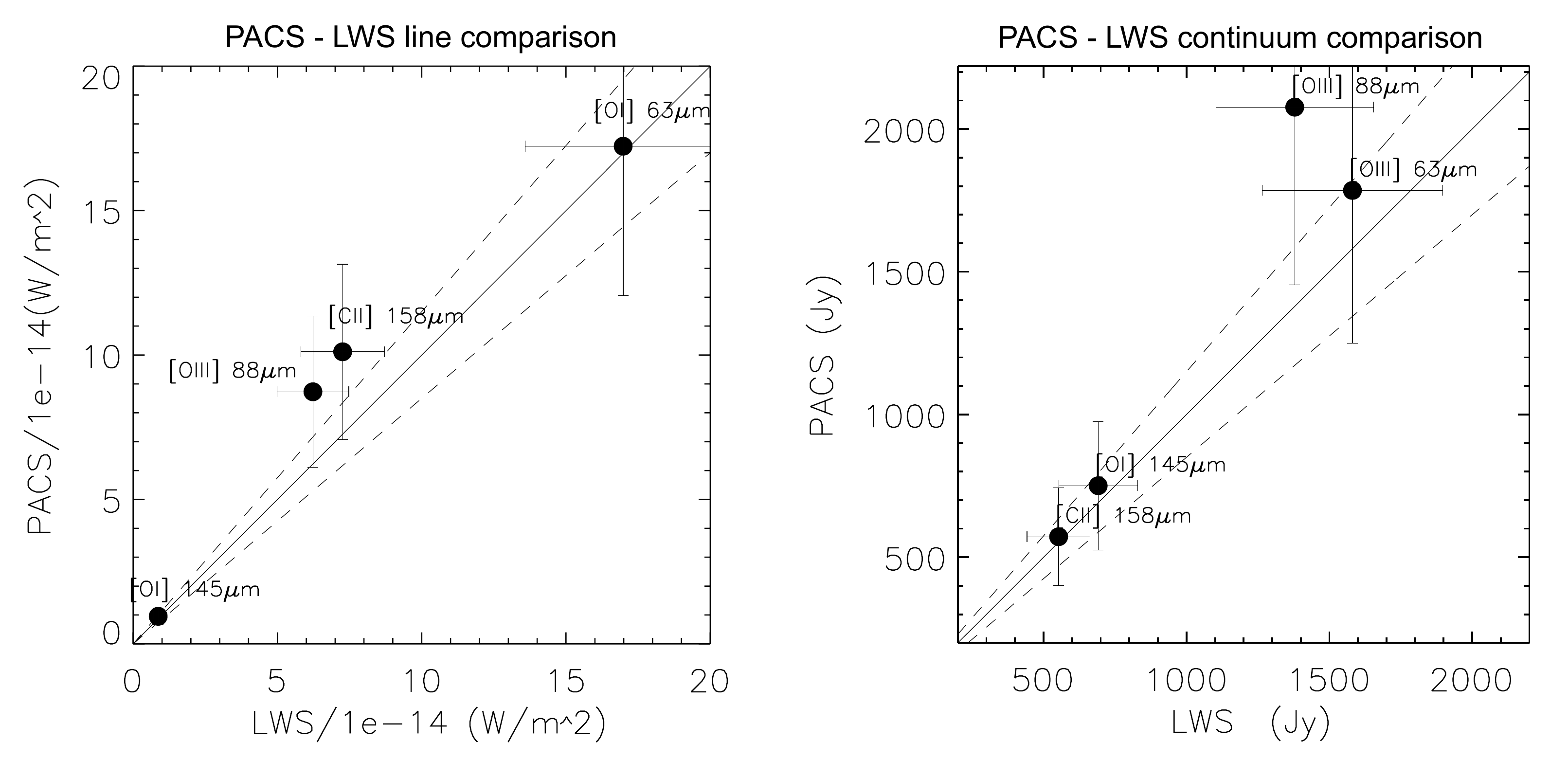}
      \caption{Comparison between the LWS and  PACS fluxes for lines (left panel) and continuum (right panel).
      The solid line indicates equal PACS and LWS fluxes, the dashed show a range of $\pm 15\%$. See text
      for details
              }
         \label{PACSLWS}
   \end{figure*}

   \begin{figure*}
   \centering
  
\includegraphics[angle=0,width=17.5cm,height=12.5cm]{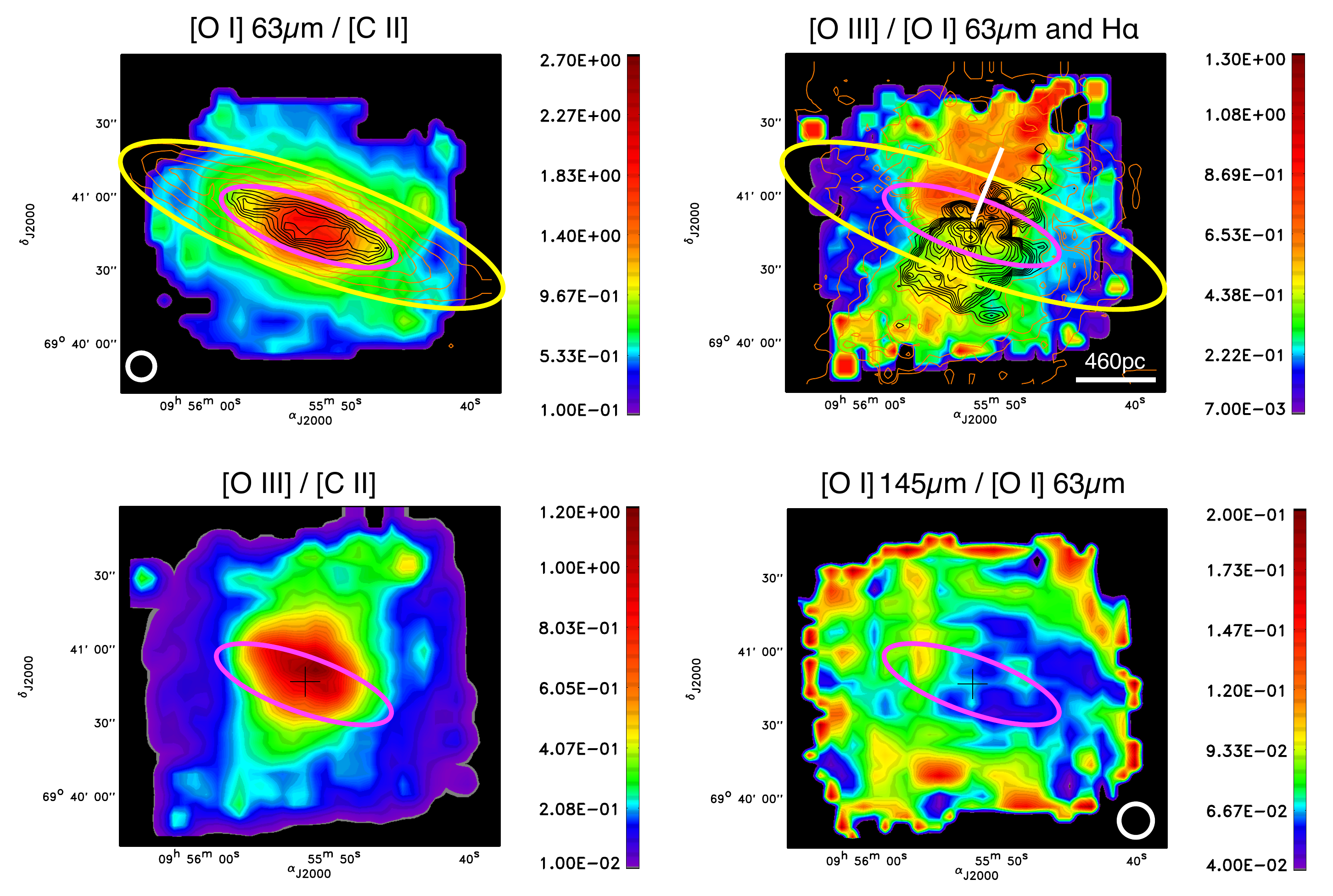}
       \caption{ [OI] 63 $\mu$m/[CII], [OIII] 88
$\mu$m /[OI] 63 $\mu$m, [OIII]  88 $\mu$m/[CII] and [OI] 145 $\mu$m/[OI] 63 $\mu$m ratio
maps. The K$_s$-band contours are overplotted on the [OI] 63 $\mu$m/[CII] ratio map. They were used to define the outer ellipse
tracing the stellar disk and the inner  ellipse that delineates  the starburst region. $H\alpha$ contours are
overplotted on  the  [OIII] 88
$\mu$m /[OI] 63 $\mu$m ratio map. The white line along the minor axis of the northern outflow, indicates a linear size of 450 pc.}
        \label{line_ratio}
   \end{figure*}

   \begin{figure*}
   \centering
  
\includegraphics[angle=0,width=17.5cm,height=6cm]{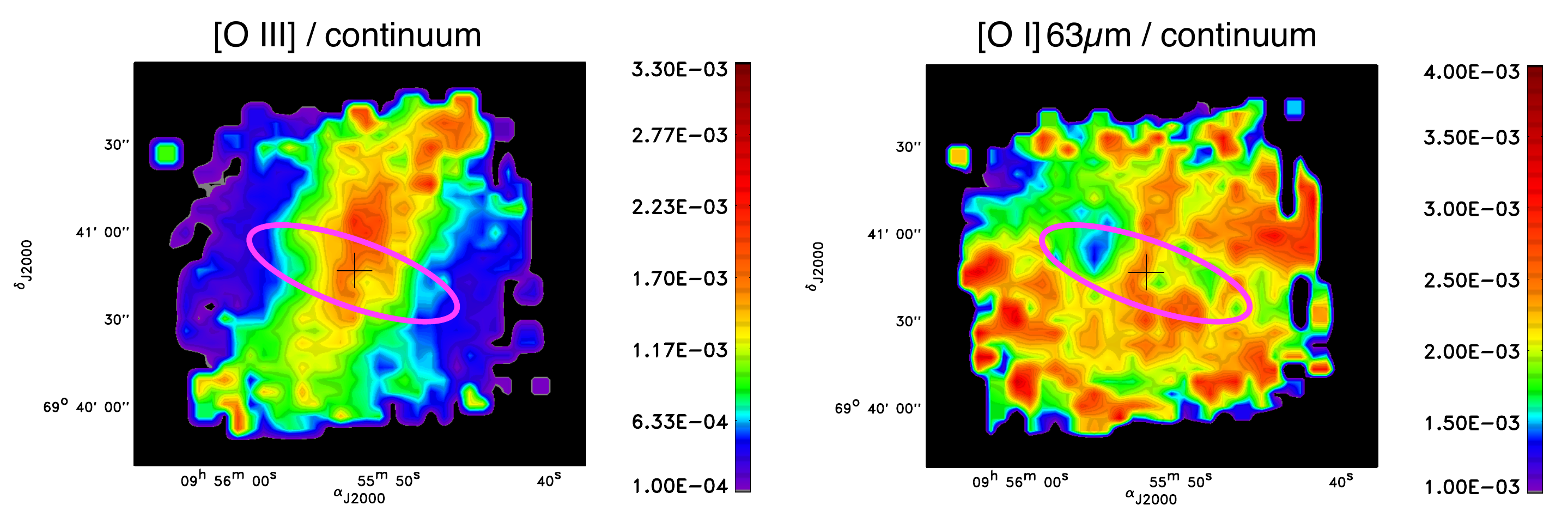}
       \caption{ [OIII] 88 $\mu$m /continuum (left) and [OI] 63 $\mu$m/continuum (right) ratio maps. The continuum is
      from the baseline level of the line fit.   }
         \label{line_over_continuum}
   \end{figure*}

\subsection{Comparison with H$\alpha$}
In order to understand where  the outflow  outlined in the [OIII]/[OI] 63 $\mu$m
ratio map is located with respect to the well known bipolar outflow of M82 
observed at other wavelengths,  the HST H$\alpha$ contours 
are overplotted on the [OIII]/[OI] 63 $\mu$m ratio map in Figure \ref{line_ratio}.
 Overall, the H$\alpha $ emission is more extended than  the  [OIII]/[OI] 63 $\mu$m ratio (orange contours 
in the upper right panel  of   Figure \ref{line_ratio}). 
The bulk of the H$\alpha $ emission, {\it {i.e.}} the "eyes" defined in Shopbell and 
Bland--Hawthorn (\cite{Shopbell}) (black contours in the upper right panel  of   Figure \ref{line_ratio}), are displaced with
respect to the region with of highest [OIII]/[OI] 63 $\mu$m ratio. Actually, the region with the brightest 
[OIII]/[OI] 63 $\mu$m ratio   almost envelopes   the  H$\alpha$ "eyes".
Also, in the northern outflow the H$\alpha$ emission is displaced to the west with respect to the 
region with the highest [OIII]/[OI] 63 $\mu$m ratio, while in the southern outflow both emission are
co-spatial.
This displacement is most likely due  to extinction of  the optical  emission; in fact,
the disk of M82 is  inclined by 81 degree in the sense that its  southern part   is offered to our
direct line of sight. Thus, the emission of the northern part of the disk and part of the
northern outflow  must suffer from a   higher  obscuration than that in southern outflow.
This agrees very well with what was found by Shopbell and Bland--Hawthorn (\cite{Shopbell}) who calculated that,
given the inclination of M82 disk, the northern outflow should be at
least partially obscured up to a projected radius of $\sim$ 48$\arcsec$. This
corresponds to the extension of the bright blob visible in the [OIII]/[OI] 63 $\mu$m
map,  north-east  from the northern H$\alpha$ emission. 

Extinction could also explain the fact that the northern outflow is detected in [OIII] at 88
$\mu$m but not at 5007 $\AA$.  
We tried to make a very rough estimate of  the extinction in the following way:
we estimated the logarithm of the H$\alpha$/H$\beta$ ratios
of both sides of the outflow  (0.6 and 1.2  for the southern and the northern respectively)  at a distance of $\sim$ 500 pc
 from the galaxy  center from  Figure 16 of Heckman, Armus and Miley (\cite{Heckman90}). With these values, 
 we could calculate 
 the  $A_V/R_V$  in the northern and  southern outflow using the Calzetti law (Calzetti, Kinney and  Storchi-Bergmann \cite{Calzetti94}).
We derived a very crude value of the observed  [OIII] 5007 $\AA$  line flux  
in the southern outflow   from Figure 2 of Shopbell and Bland--Hawthorn (\cite{Shopbell}) and, using a slab geometry, the 
corresponding dereddened flux. Assuming that the intrinsic [OIII] 5007 $\AA$ emission in the northern outflow is 
equal to that of the southern outflow, we estimated the expected flux, using the $A_V/R_V$ values obtained 
for the northern outflow.
We find that the expected [OIII] 5007 $\AA$ flux is $\sim 2\times 10^{-17}~ \rm{erg~s^{-1}~\rm{cm^{-2}}}$ for the slab model 
which is appropriate for the Calzetti law.
 This flux is smaller than the minimum flux reported by Shopbell and Bland--Hawthorn
 (\cite{Shopbell}) in their Figure 2,
and it is compatible with the non-detection of the  [OIII] 5007 $\AA$ line in the northern outflow. We stress, however, that 
this calculation is highly uncertain, first, because of the very uncertain input values which we had to deduce from
 the published  figures, second because there is no physical reason to assume that the intrinsic [OIII] 5007 $\AA$ emission in the two cones of
 the outflow should be the
same.

 One important question is whether in the outflow, the bulk of [OIII] 88 $\mu$m line emitting gas is associated with the bulk 
of the ionized gas emitting 
emitting in H$\alpha$. 
As  already mentioned,  these two emission components show different kinematics, with the [OIII] 88 $\mu$m emitting gas
being slower ($\sim75 ~\rm{km~s^{-1}}$) than  the H$\alpha$ emitting gas ($\sim 600~ \rm{km~s^{-1}}$, Shopbell and Bland--Hawthorn  \cite{Shopbell}).
We have also seen that  the [OIII] 88 $\mu$m emitting gas does not show  any line splitting, contrary 
to what has been  observed in the outflowing gas emitting in H$\alpha$. The H$\alpha$ line splitting in the outflow
of M82 has been interpreted as due to the confinement of the  H$\alpha$ emitting gas  in  the  walls of the outflow (Greve et al. \cite{Greve}). 
These facts suggest a possible different spatial distribution of the ionized gas traced by the H$\alpha$ and 
the [OIII] 88 $\mu$m lines: while the first is confined in the outflow walls,
the second is  more  concentrated along the minor axis of the galaxy.\\
This  may also suggest that these  gas components  are   ionized by two different physical processes.
In fact, in the commonly accepted  scenario, the bulk of the H$\alpha$ emitting gas is confined in the walls of the outflow,
because it arises, at least in part,  from  shocked gas in the region of 
interaction between the X-ray emitting gas and the halo. This naturally explains the observed 
 H$\alpha$ line splitting.  Instead, the [OIII] 88 $\mu$m line may arise  
principally from ionization due to the starburst photons.
Although we do not have enough data available to prove that this is the case, previous results seem to support this scenario.
Shopbell and Bland--Hawthorn
 (\cite{Shopbell}) show that in the southern outflow, the optical [NII]/H$\alpha$ ratio is lower along the minor axis  
 than further away from the axis, probably reflecting   a higher ionization. This is interpreted as radiation 
 produced by the two star forming  clusters in the disk escaping along the minor axis of the disk. 
Veilleux, Rupke and Swaters (\cite{H2}) reached  a similar conclusion for both
parts of the outflow studying the  
variation of the ratio between the near-infrared $H_2$  and the   PAH emission.

We can summarize the main results obtained in this Section as follows.
{\it {Both neutral and ionized atomic gas  are detected in the outflow and  in the disk. 
In the outflow, the ionized gas  emitting in the [OIII] 88 $\mu$m line   is more
collimated  than the neutral gas and  also than the ionized gas emitting in H$\alpha$, which is more confined in the 
walls of the biconical outflow. The ionized 
gas traced by the   [OIII] 88 $\mu$m  line and the neutral atomic gas  show the same kinematics in
the disk and in the outflow which differ  from the kinematic  of the H$\alpha$ emitting gas. 
These facts suggest a scenario where, while  the H$\alpha$ filaments  in the outflow   arise, at least in part, from shocked gas
in the regions of interaction between  the outflowing hot plasma emitting in X-ray and the halo of the galaxy, the
bulk of the [OIII] 88 $\mu$m line emission arises from gas photoionized by the starburst radiation along the galaxy minor axis.
 The comparison between the optical and infrared   emission
suggests that the base of the northern outflow is more extincted than the southern one, probably obscured by
the disk itself.}}  \\

\section{Analysis}
In this section, we will address the origin of the neutral material detected with PACS, with particular focus on  its origin in
the outflow. We have seen that this is certainly material contributing to the outflow and the question is whether this
emission  can be explained by classical Photo--Dissociation-Regions (PDRs) emission and, if this is the case, 
how  the physical conditions of the PDRs 
in the outflow and  in the galaxy disk  compare to each other. We will also compare the energetics and the kinematics of the neutral medium with
 that of the other gas phases detected in the outflow of M82.\\

\subsection {Photo-dominated regions  diagnostics}
The [CII] and [OI] lines are the main coolants of the atomic medium. They are collisionally excited 
by  the atomic gas heated from electrons ejected from grains  via  photoelectric effects.
Since C$^+$ has an ionization potential equal to 11.3 eV, it is mainly excited by Far UV (FUV)  photons
(h$\nu$$<$ 13.6 eV) escaping from HII regions (but it can also be excited  in the HII regions 
themselves as we will discuss in Section 4.1.3).
The regions where  FUV photons dominate the ISM physics are called Photondominated or Photodissociation Regions (PDRs, 
Hollenbach and Tielens \cite{Hollenbach99} and reference therein). Since the atomic gas cooling is mainly traced 
by the [CII] and [OI] line at 63 $\mu$m  and the heating is due to the
photoelectric effect on the grains,  the
ratio between the total  emission from the FIR fine structure lines and the
total  infrared (TIR) emission from grains
gives the total  photoelectric yield.  Comparing  the observed
yield with that predicted by the model one can derive fundamental  physical  
parameters of the neutral  ISM associated with the PDRs.

In this  section, we will show the results obtained by applying the PDR
model  from Kaufman et al. (\cite{Kaufman}) to the PACS data of M82. 
By applying the model to each pixel of our maps, {\it {we are able
to produce,  for the first time for an extragalactic object, high resolution, fully sampled maps of the FUV interstellar
radiation field (ISRF) expressed in G$_0$ units{\footnote{  G$_0$ is the FUV (6--13.6 eV) ISRF 
normalized to  the solar neighborhood value expressed in Habing flux:
1.6$\times$10$^{-3}  \rm{erg~ s^{-1} cm^{-2}}$}}, of   the neutral gas density n$_H$, of  the gas temperature 
at the surface of the modeled clouds T$_S$ and  of the cloud beam filling factors
$\phi$.}}
Because of different step sizes in mapping among all wavelengths and final  dimensions of the various continuum
maps, we  rebinned all maps to the  grid used for 
the longer wavelengths mapping (6$\arcsec$/pix),   reduced all images to the same size,
 smoothed all maps to the coarser spatial  resolution of the [CII] map, then aligned all to the [CII] emission peak.
In the following analysis we will assume that the [CII] and [OI] lines arise from PDRs also in the outflow, 
and we will check {\it a posteriori} if this makes physically sense.
However, in order to correctly and successfully apply the PDR diagnostics 
one has  to make some corrections  to the observed fluxes that we will explain in detail in the following sections, 
 before showing the results.

\subsubsection{Total infrared emission map calculation}
One fundamental input to the PDR modeling is the total IR emission (TIR from $\sim 30$ to $1000~ \rm{\mu m}$). 
Since our aim is to derive a solution for every pixel of the M82 mapped area, we need to have also the TIR emission
in each pixel. 
 But the IR continuum data available from previous observations (IRAS, ISO, SPITZER) 
 do not have the spatial resolution that matches that of PACS at these wavelengths.
Fortunately, thanks to the brightness of M82, we could use the continuum detected aside the  targeted
lines and in the corresponding parallel channel, as photometric points   from which we could
derive the FIR ( from $\sim 40$ to $200~ \rm{\mu m}$) emission at each pixel.
 This gives a total of   8 points per pixel at 
63, 72.8, 78.9, 145, 157, 176.8 and 189.6 $\mu$m.
We fit  the continuum fluxes with a modified blackbody, in the 
 optically thin emission approximation and  with a fixed value of
the grain emissivity, {\it i.e.} with $\beta$ = 1
in  each pixel. However, the agreement of the set of continuum data points obtained from the 
blue  band lines and the corresponding red parallel points do not agree very well with those
obtained from the continuum aside the red lines and the corresponding blue parallel data.
In particular, the data derived from the parallel channels of the [CII] and [OI] line at 145 $\mu$m ({\it i.e.} 72.8 and  78.9 $\mu$m) 
are often systematically higher than the  data values at the other wavelengths. 
The reason for this is not well understood, but it might correspond to an higher continuum distortion at these wavelengths introduced 
from the bright line mode with respect to other wavelengths which increases the uncertainties.\\
Therefore we restrict the  fitting to the 4 photometric points
calculated on the blue lines  and the corresponding parallel data. 
In this way we include in the gray body fitting the point at the highest possible wavelengths,
allowing a better determination of the far infrared emission. 
An example of such a fit is shown in Figure \ref{bb}  for one pixel.
From the black body fit,  for each pixel we calculated the flux density at 60 and 100 $\mu$m which 
is necessary for the comparison  with diagnostics used in the pre-Herschel era, and from
these flux densities we derived the  FIR (from 42.5 to 122.5 $\mu$m)  emission  using 
the Helou (1988) formula :
\begin{equation}
FIR (W/ m ^{2} )=1.26 \times 10 ^{-14} \times (2.58 \times I _{60 \mu m} (Jy)+
I _{100 \mu m} (Jy)).
\end{equation}

The blackbody fitting also results in a dust temperature for each pixel that we used to
build the map  shown in Figure \ref{Tdust} where also the $H\alpha$ contours and the smaller of the two ellipses
 shown in previous  Figures
are shown.  Taking into account the flux uncertainties 
we estimated an uncertainty on the dust  temperature of $\pm$ 3 K.
The dust temperature image  shows very clearly the bright
central  region included in the ellipse and corresponding to the brightest $H\alpha$ emission. 
Here the dust temperature ranges from $\sim$ 45 to    $\sim$ 55 $K$,
 typical for a starburst region.
In the map  is also visible an extention of the dust emission  above the disk along the minor axis
 which coincides with the H$\alpha$ extension along this direction and with  the outflow seen in 
 the [OIII]/[OI] 63 $\mu$m ratio map.
The dust in the southern outflow  is slightly  hotter than in the northern outflow.\\
From the FIR emission we calculate the TIR emission needed as  input to the Kaufman et al. (\cite{Kaufman}) model 
 by   using   eq. 3  given in
 Dale et al. (\cite{Dale01}) which calculates the conversion from FIR to TIR from the 60/100 $\mu$m ratio.

   \begin{figure}
   \centering
\includegraphics[angle=0,width=8.5cm,height=7cm]{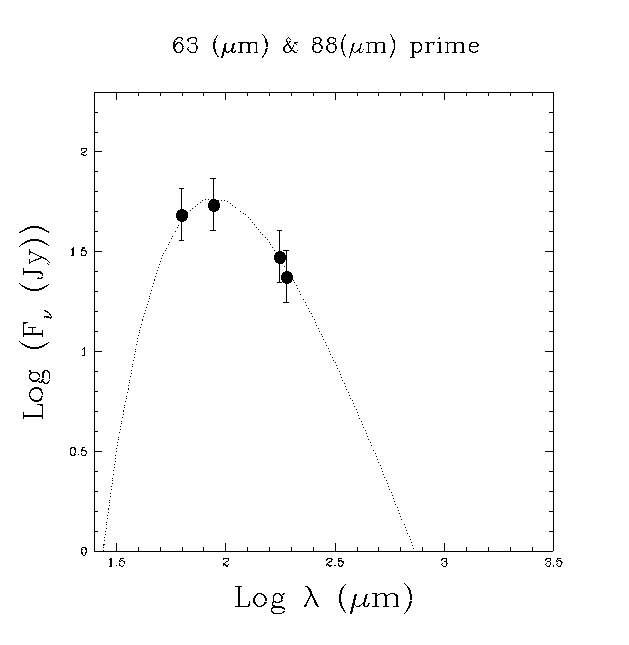}
       \caption{Example of a modified blackbody fit on one pixel in the center of M82. The points correspond  to
        the continuum averaged detected with PACS on both sides of the [OI] line at 63 $\mu$m and [OIII] at 88 $\mu$m and on the corresponding parallel channel.}
         \label{bb}
   \end{figure}

\subsubsection{Geometrical effect correction factor.}
The PDR model we consider in this
work  assumes the emission is generated by a  one-dimensional semi-infinite slab illuminated from one side. The model also considers [CII], FIR and [OI] 
at 145 $\mu$m optically thin, and [OI] at 63 $\mu$m optically thick,  such that in a model where  a  cloud
is illuminated from one side, only one side of the cloud emits in the [OI] line at 63 $\mu$m while [CII], FIR and [OI] at 145 $\mu$m  emits
from both the front and the back of the  cloud.  At the PACS resolution each beam includes many clouds, and the
orientation of the illuminated side will be randomly distributed. This is equivalent to consider that the beam includes one cloud illuminated
from all sides. The model would then predict half of the [OI] at 63 $\mu$m. This is why we multiply the observed
[OI] at 63 $\mu$m by 2 before applying the PDR models.  Since the [OI] at 145 $\mu$m is considered to be thin in the model, we do not have
to apply any correction to it.

\subsubsection{Contribution to the [CII] emission from the ionized gas}
 [CII] can also arise from the ionized medium. Since the PDR model 
considers  [CII] arising only from the neutral medium, before applying the model to the
observations we have to subtract this contribution from the observed [CII] emission.
The percentage of the [CII] emission arising from ionized gas with respect to
the total observed emission,  is minimum ( $\sim 10\%$) in dense HII regions, since here
most of the carbon is double ionized, but it increases significantly  in the diffuse ionized medium,
the precise amount depending on the physical parameters of the medium (Nagao et al. \cite{Nagao}).
In external galaxies the beam typically encompasses regions with very different physical conditions, 
therefore one has to guess  the average properties of the observed medium in each spatially 
resolved region, in order to estimate the [CII] contribution from ionized gas.
This will  obviously introduce some uncertainty in the derived parameters.

The most straightforward way to determine the fraction of [CII] arising from the ionized medium 
is to use  the  scaling relation between the  ($^3$P$_1$$\rightarrow$$^3$P$_0$)  [NII] line at 205.178 $\mu$m and the [CII] line 
(Abel \cite{Abel}).
Since these have  similar critical electron densities, (80 and 50 $\rm{cm^{-3}}$ respectively), their ratio is almost
independent on the electron density.
Unfortunately,  we cannot use this method because we did not observe  M82 in the [NII] 205 $\mu$m line with PACS.
A less direct, and more uncertain, way  to calculate the  contribution from ionized gas to the  observed [CII] line
flux,  is to use   the  [NII]  122 $\mu$m line. Since [NII]  122 $\mu$m has a higher critical electron density  
(3.1$\times$10$^2$  $\rm{cm^{-3}}$) than that of the [CII] line, the  ratio of these lines depends  strongly on the density. 

   \begin{figure}
   \centering
\includegraphics[angle=0,width=9.1cm,height=6.5cm]{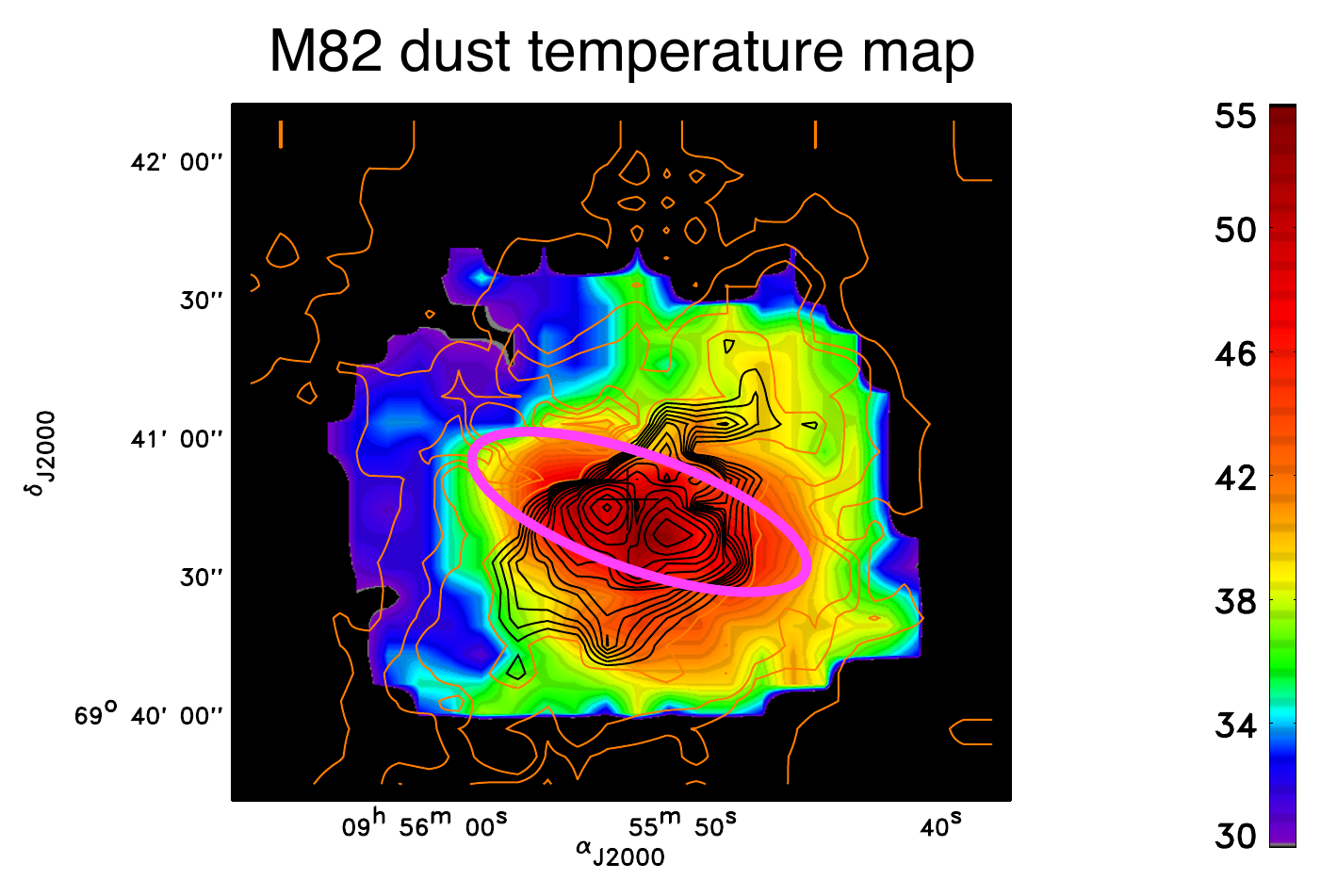}
       \caption{Dust temperature map. Units are $^\circ$K. H$\alpha$ contours are overplotted. Note that these contours are at a resolution
       poorer than what  is shown in Figure \ref{line_ratio} to match the pixel size of the dust temperature map.}
         \label{Tdust}
   \end{figure}
 
We have observed M82 in the  [NII]  122 $\mu$m  line with three PACS spectrometer pointings centered on the
disk, and on the north and south outflow as shown in Figure \ref{Staring}. 
We have divided the galaxy in four regions: the starburst, the north and the south outflow  and the
more diffuse disk and we have calculated the mean [NII] emission for each of them 
using the average of the spectra belonging to each of the defined regions. 
The resulting average [NII] intensities in the starburst, disk, and in the northern and southern parts of the 
outflow  are  87.4, 1.0,
 5.3 and 3.4 $\times$ 10$^{-17}$   $ \rm{W ~ m^{-2}}$.
We assume that in the disk and in the starburst region of M82 the [CII] fraction arising from ionized gas scales with the observed [NII] following the
relation valid for the  dense HII regions and  for a  Milky Way C/N  abundance ratio equal to 3.8, Rubin et al.
(\cite{Rubin88}, \cite{Rubin93}):

\begin{equation} 
[CII]_{ionized}^{Disk}=[CII]_{dense~HII}^{Disk}=1.1 \times [NII]^{Disk} .
\end{equation}

In a recent work, Croxall  et al. (\cite{Croxall}), show how the [CII]/[NII] 122 ratio varies with electron
density (their Figure 11). We do not have a direct electron density measure in the outflow  but we know from previous
works that the average electron density of the   ionized gas traced by the H$\alpha$ is $\sim 15~\rm{cm^{-3}}$  (Shopbell and Bland-Hawthorn
   \cite{Shopbell})   and   the estimated averaged electron density in the outflow  is
$\lesssim 300~ \rm{cm^{-3}}$ (Yoshida, Kawabata and Ohyama \cite{Yoshida}). Since  we  do not know
{\it a priori} from which of these these media the [CII]  arises, we have calculated the amount of [CII] arising from
ionized gas for these two cases by using the ratio published by Croxall et al.  (\cite{Croxall}). 
From this work we derive a   [CII]/[NII] 122 ratio equal to 3.7 and 0.7 for $n_e = 15$ and $n_e = 300~
\rm{cm^{-3}}$. Note that  these two density values correspond
more or less the minimum and the maximum correction possible, in order not to
have negative [CII] results.
 In each case, the  resulting [CII] fluxes were then subtracted from the observed [CII] map, in the region defined 
 as outflow.  
The resulting percentages of the [CII] emission  arising from the ionized gas  are  $\sim$ 10, 30  $\%$ in the disk 
and in the starburst,  and $\sim$ 11 (47) $\%$  and   $\sim$ 11 (40) $\%$
in the north and south cones of the outflow  based on $n_e= 300~(15)~ \rm{cm^{-3}}$ respectively. \\

We stress, however, that these estimates of the  contribution to the observed [CII] flux from pure ionized medium  
 are very uncertain especially when applied to regions with an unknown and complex mix of different ISM phases. 
Moreover, because of the lack of a fully sampled map in the [NII] 122 $\mu$m line map, we were
forced to use an average value of the [NII] 122 emission for each of the four macro-regions 
we have defined in  the galaxy.   
 It is for this reason  that we decided to  run the model also
with and without the corrections to the observed [CII] emission line we have explained above, 
to check whether the main results we derive  from the modeling  are  independent of the correction applied. \\
Moreover, in principle  we could have used as input for the PDR modeling the two [OI] lines, [CII] and TIR. 
We have tried this but the uncertainties
in each of the line is such that the result is much nosier  than using 3 input values  with only one [OI] line at the time. Therefore, we run the model
using always only three input values, {\it i.e} the [CII], the FIR and one of the [OI] lines  at the time,  and  with and without the ionized 
contribution correction to the [CII] emission line ({\it i.e.}  6 input sets in total).

\begin{table}
\caption{ Main PDR physical parameters obtained by modeling the [CII], [OI] 145 and FIR emission with and without 
correcting the observed [CII]  line flux from the contribution from various ionized gas (see Section 4.1.3 for details). 
The listed  values are  calculated
 by averaging the maps shown in figures from \ref{PDRsol_G0} to \ref{PDRsol_phi}, over the four
macro regions we have divided the galaxy in: starburst, disk, north and south outflow (see figure \ref{Gnplot} for a 
schematic visualization of these regions). Last column lists the averaged value for the [CII] opacity maps obtained as explained in Section 4.3. }              
\label{Table1}      
\centering                                      
\begin{tabular}{c c c c c c c c}          
\hline   
 [CII] correction applied           & $G_0$              & $n_{H}$         & $T_{gas}$        &     $\phi$          &   $\tau_{CII}$     &   \\
                                    &                    & $ \rm{cm^{-3}}$ &  K               &       $\%$              &        &  \\
\hline   
                                    &                    &                 &                  &                     &        &    \\
{\bf Starburst}                     &                    &                 &                  &                     &        &    \\
No correction                       &     1400           &     760 	   &	  460         &       6.5 	    &   0.06     &    \\   
Dense HII region                    &     2500           &     900 	   &	  470         &       5.5 	    &   0.03      &    \\   
\hline   
                                    &                    &                 &                  &                     &        &     \\
{\bf Disk}                          &                    &                 &                  &                     &        &     \\
No correction                       &     400 	         &     200	   &      380 	      &	      1.6           &   0.1     &    \\   
Dense HII region                    &     450 	         &     220 	   &      390 	      &       2.4           &   0.05     &    \\   
\hline   
                                    &                    &                 &                  &                     &        &       \\
{\bf North Outflow}                &                    &                 &                  &                      &        &      \\
No correction                       &     340	         &     380 	   &	 300	      &       2.3 	    &  0.03      &     \\   
$\rm{n_e} = 15$ $\rm{cm^{-3}}$      &     960 	         &     600 	   &	 390 	      &       1.5 	    &  0.02      &     \\   
$\rm{n_e} = 300$ $\rm{cm^{-3}}$     &     430            &     450  	   &	 320	      &       1.7 	    &  0.03      &     \\   
\hline   
                                    &                    &                 &                  &                     &        &    \\
{\bf South Outflow}                &                    &                 &                  &                      &        &    \\
No correction                       &     350	         &     210 	   &	 350 	      &	      1.0 	    &   0.06	 &    \\   
$\rm{n_e} = 15$ $\rm{cm^{-3}}$      &    1200 	         &     420 	   &	 440 	      &	      1.2 	    &   0.03	 &    \\   
$\rm{n_e} = 300$ $\rm{cm^{-3}}$     &     390	         &     210 	   &	 420 	      &	      1.6 	    &   0.04     &	  \\   
\hline   

\end{tabular}
\end{table}

\subsection{Results from the PDR modeling}

By applying the  PDR model  for each input data set, we obtained  $G_0$ and n$_H$ for each pixel where all three input values are
detected. From $G_0$ and n$_H$  we then derived the corresponding
 gas temperature at the surface of the modeled cloud ($T_S$)
 and the area beam filling factor ($\phi$), {\it i.e.} the percentage of the beam covered by the emitting
 clouds, 
 obtained by  dividing the [CII]
 emission corresponding to the $G_0$ and n$_H$ solutions and the input [CII] value.
Since this can be done for each pixel we were able to obtain 6 such maps for each input
 dataset defined as explained in the previous section.
 We stress,  however, that given the flux uncertainties and the fact that a model is an oversimplification 
 of much more complex real physical conditions, we estimate the
 uncertainties of the derived PDR model parameters to be a factor $\sim$ 2. Also we do not really 
 trust the pixel-to-pixel
  variation in the PDR solution maps, simply because the model uses a discrete grid, and given the flux uncertainties 
  it is possible 
  that a resulting solution is higher or lower than  the solution it would have been  assigned with smaller uncertainties.
  Sometimes this is visible in a hot/cold pixel in a middle of a well defined morphological structure, and we corrected for these
  interpolating with the neighborhood pixels.
 
Figures   \ref{PDRsol_G0} to \ref{PDRsol_phi} show   the   6   maps obtained using the   six  sets of inputs,   for each of the above
 mentioned physical parameters.
 In detail:  the top panels show the solutions obtained  using as input to the PDR modeling   [OI] 63, [CII] and FIR. 
 Two  of these maps were obtained by   correcting the observed
 [CII] emission from  the contribution arising from ionized gas  assuming  electron  densities equal to 300 
 and 15 $\rm{cm^{-3}}$ in the outflow  and applying equation 2   in the disk and starburst regions. The third solution 
   was obtained without correcting   the observed [CII].  
 The bottom panels of figures \ref{PDRsol_G0} to\ref{PDRsol_phi} show the three sets of solutions   
 obtained in a similar way but using as input to the model the [OI] line at 145 $\mu$m instead of the [OI] line at 63 $\mu$m.
  
 All maps are very similar in morphology except for  the gas  density maps 
 (Figure \ref{PDRsol_nH}). In particular, this figure shows that the gas
 density is  enhanced in the starburst region for all the solutions obtained   using  
  the [OI] line at 63 $\mu$m, and in the  
  northern part of  the starburst
 when the [OI] line at 145 $\mu$m is used. \\
   We find that this difference is most likely due to misalignment 
 between the [OI] 63 $\mu$m and [CII] images. We explain this in details in the Appendix. 
 Once this misalignment has been corrected, 
 the morphologies of the  images obtained by the PDR modeling for all parameters, included the density $n_H$,   agree (see Figure \ref{PDRsol_nH2}),
 and the values  agree within a factor of $\sim$ two for all maps, except  the density map for which the biggest difference is close to  
 a factor of $\sim$ 3. \\
Because of the problems and uncertainties discussed above, and the fact that the [OI] 63 $\mu$m  optical depth can
significantly vary in the galaxy,    we will use  for the following analysis 
 only the   results obtained by using   as input
 the [OI] 145 $\mu$m line. This  reduces the
 uncertainties of our analysis   to those related to the    [CII] correction for ionized gas,
  since the [OI] 145 $\mu$m line is optically thin. Moreover, the pointing variation and the the processing  
 affects  the [CII] and [OI] 145 $\mu$m maps in a similar way, since they have been observed in the same AOR.  
 The   main uncertainty left, {\it i.e.}   that   due to
     the [CII] correction for the ionized gas contribution,  can be assessed  by examining the results 
obtained from the   corrected and the not corrected [CII] values.

 Therefore, we now concentrate on the results shown in the bottom panels of figures   \ref{PDRsol_G0} to \ref{PDRsol_phi}.
 The first thing to notice  is that, for each of the derived maps,  there is no significant difference in the morphology 
 among the solutions obtained with no [CII] correction, and those with the two different corrections we have applied 
 to the observed [CII] flux. We have calculated the average values of $G_0$, $n_H$, $T_{gas}$ and 
 $\phi$  in the four regions we have considered for the different correction to the observed [CII], for each case, and listed them in
 Table~\ref{Table1}.
   The ranges of values of 
 the PDR solution parameters are all comparable within the 
uncertainties (a factor of 2$\sim$3).  We are thus confident that the   correction we have applied to the observed [CII] does not
really play a major role. \\
 As already mentioned, the gas density maps, show  
  a density enhancement in the upper part of the starburst region, corresponding also to the brightest 
  emission in the
 [OIII]/[CII] ratio map  shown in 
 Figure  \ref{line_ratio}.\\
 The gas temperature is everywhere greater than $\sim 300$ K as we had already deduced from the constancy 
 of the two [OI] lines ratio map. The brightest emission outlines the disk of the galaxy and no structures 
 corresponding to the outflow are visible.\\
The PDR area filling factor maps are very similar in all cases and show that there is an enhancement in the 
galaxy disk with a slight asymmetry toward its northwestern part. The area filling factors in the outflow 
are significantly smaller than in the
 disk and very similar to the diffuse more spherically distributed  emission visible in all line intensity maps.  \\
  One interesting result common to all solutions is that we mapped an area sufficiently large to detect relatively diffuse emission from the 
  galaxy disk, with $G_0 \sim 100$ and $n_H \sim 50~ \rm{cm^{-3}}$. Previous space missions have not had the sensitivity 
  or spatial resolution to detect FIR fine structure lines from the low density cold atomic medium of galaxies.
   As demonstrated by our results, Herschel will greatly advance the study of the multi-phase ISM in external galaxies in the FIR.

   \begin{figure}
   \centering
\includegraphics[angle=0,width=7.5cm,height=7cm]{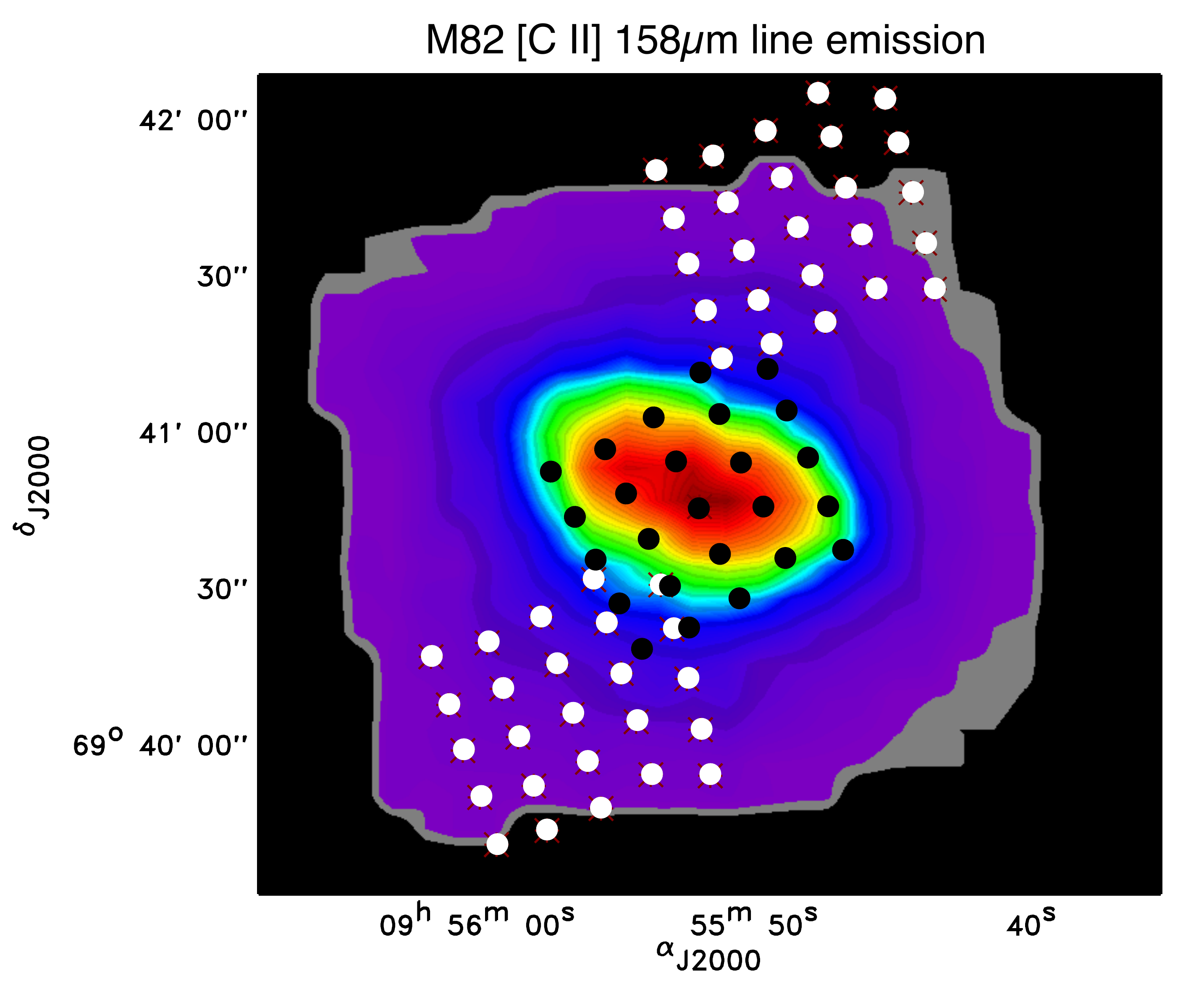}
       \caption{The PACS spectrometer footprints of the 3 pointings executed  in [NII] 122 $\mu$m are
       overplotted on the [CII] intensity  map of M82 (the same as shown in Figure \ref{line_int}). Note that the
       size of the circles does not represent the size of a spaxel on the sky.}
         \label{Staring}
   \end{figure}

   \begin{figure*}
   \centering
\includegraphics[angle=0,width=19cm,height=11.0cm]{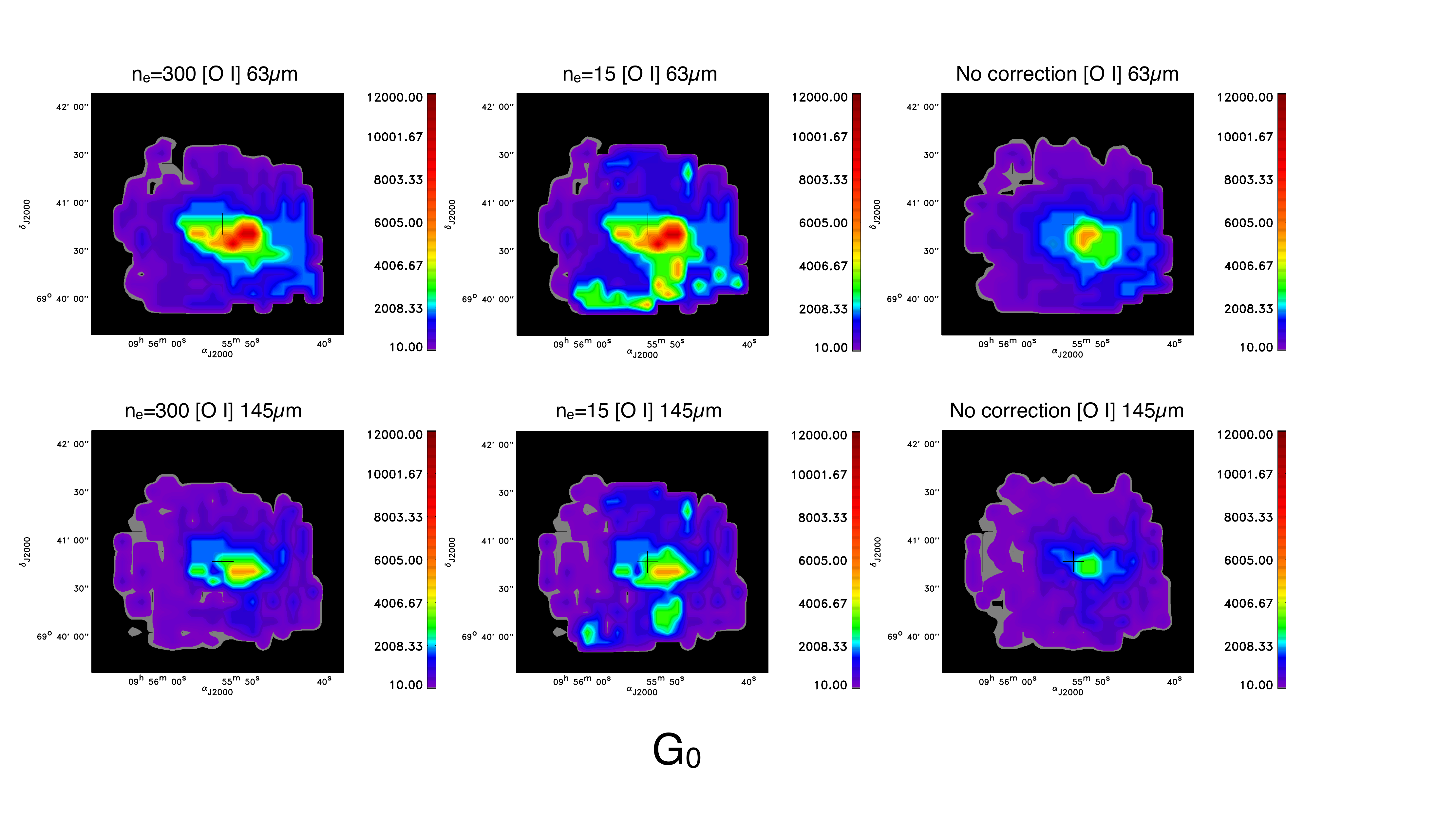}
       \caption{The  PDR solution maps obtained for  the  Far UV Interstellar Radiation Field in units of $G_0$.  
       Left to right are the results obtained correcting the [CII] flux from the ionized contribution obtained by
        considering the densities of the ionized gas  equal  to 300 (left) and 15 $\rm{cm^{-3}}$ (middle)  and 
	with no correction (right). The top raw shows
        the results obtained using as input   [OI] 63, [CII] and FIR; the bottom raw shows the results obtained
        using as input [OI] 145, [CII] and FIR (see Section 4.3.1 for details).     }
         \label{PDRsol_G0}
   \end{figure*}

   \begin{figure*}
   \centering
\includegraphics[angle=0,width=19cm,height=11.0cm]{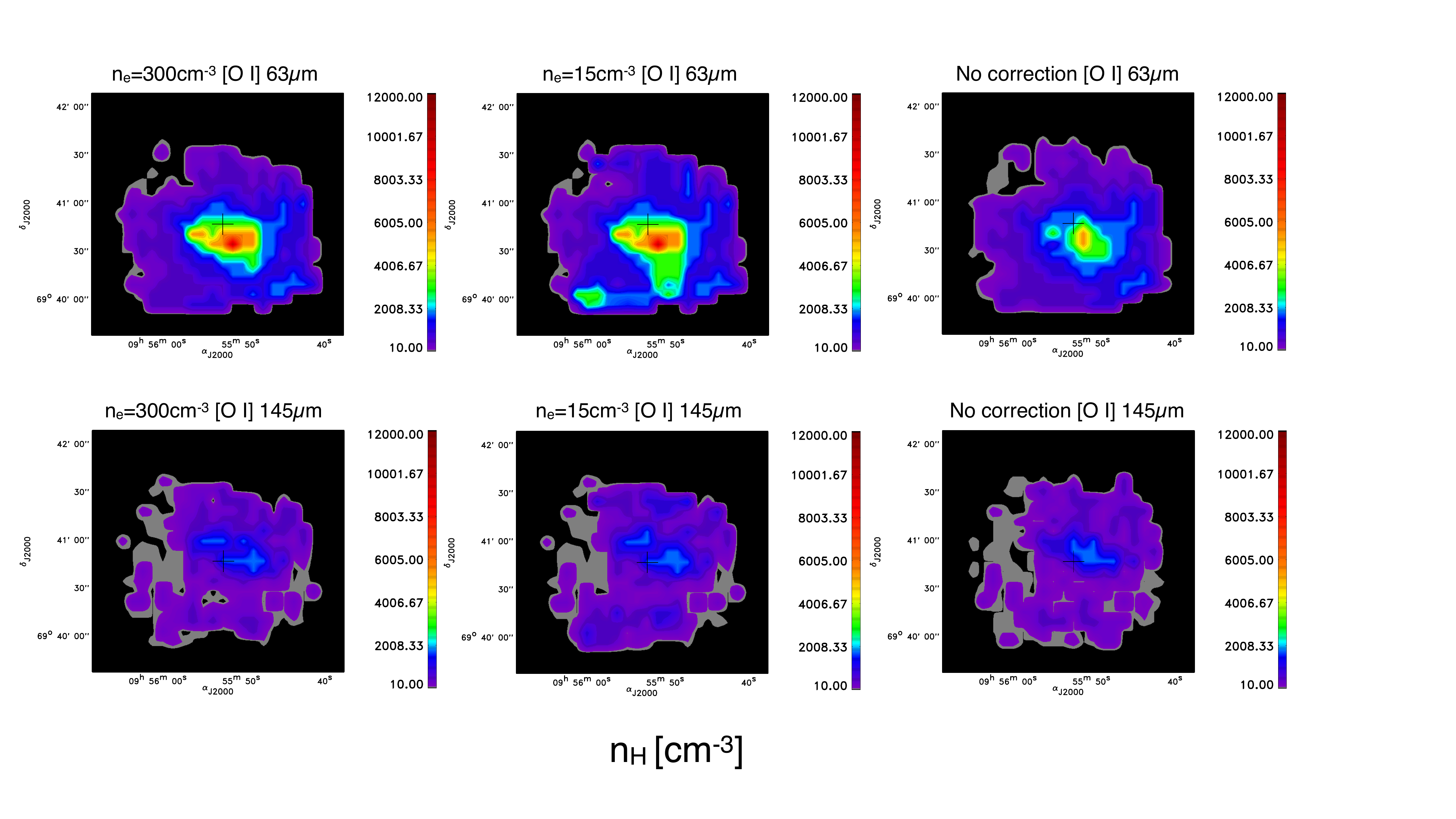}
       \caption{The  PDR solution maps obtained for the gas density $n_H$. }
         \label{PDRsol_nH}
   \end{figure*}

   \begin{figure*}
   \centering
\includegraphics[angle=0,width=19cm,height=11.0cm]{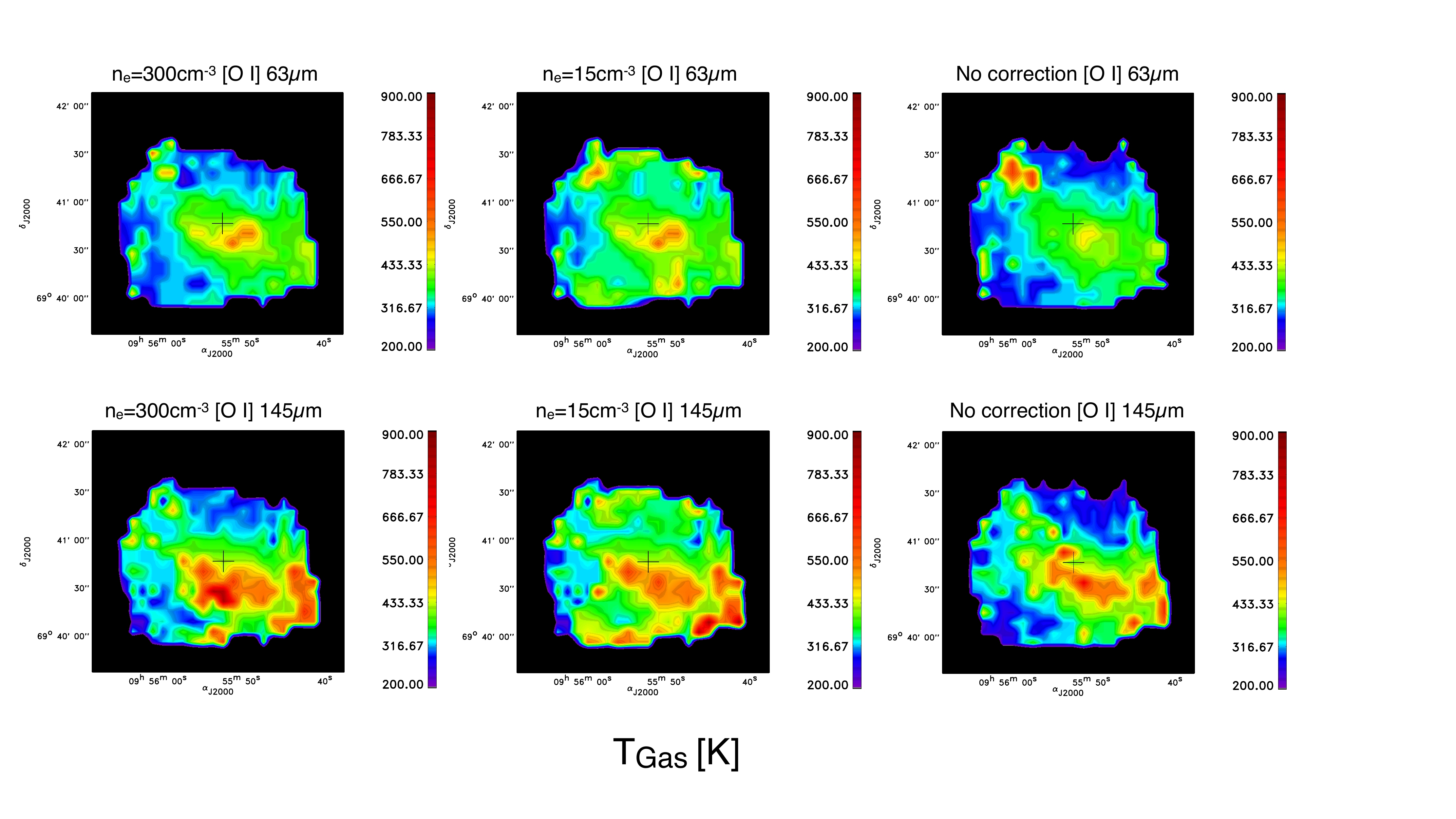}
       \caption{The  PDR solution maps obtained for  the temperature of the gas $T_{gas}$}
         \label{PDRsol_Tgas}
   \end{figure*}
 
   \begin{figure*}
   \centering
\includegraphics[angle=0,width=19cm,height=11.0cm]{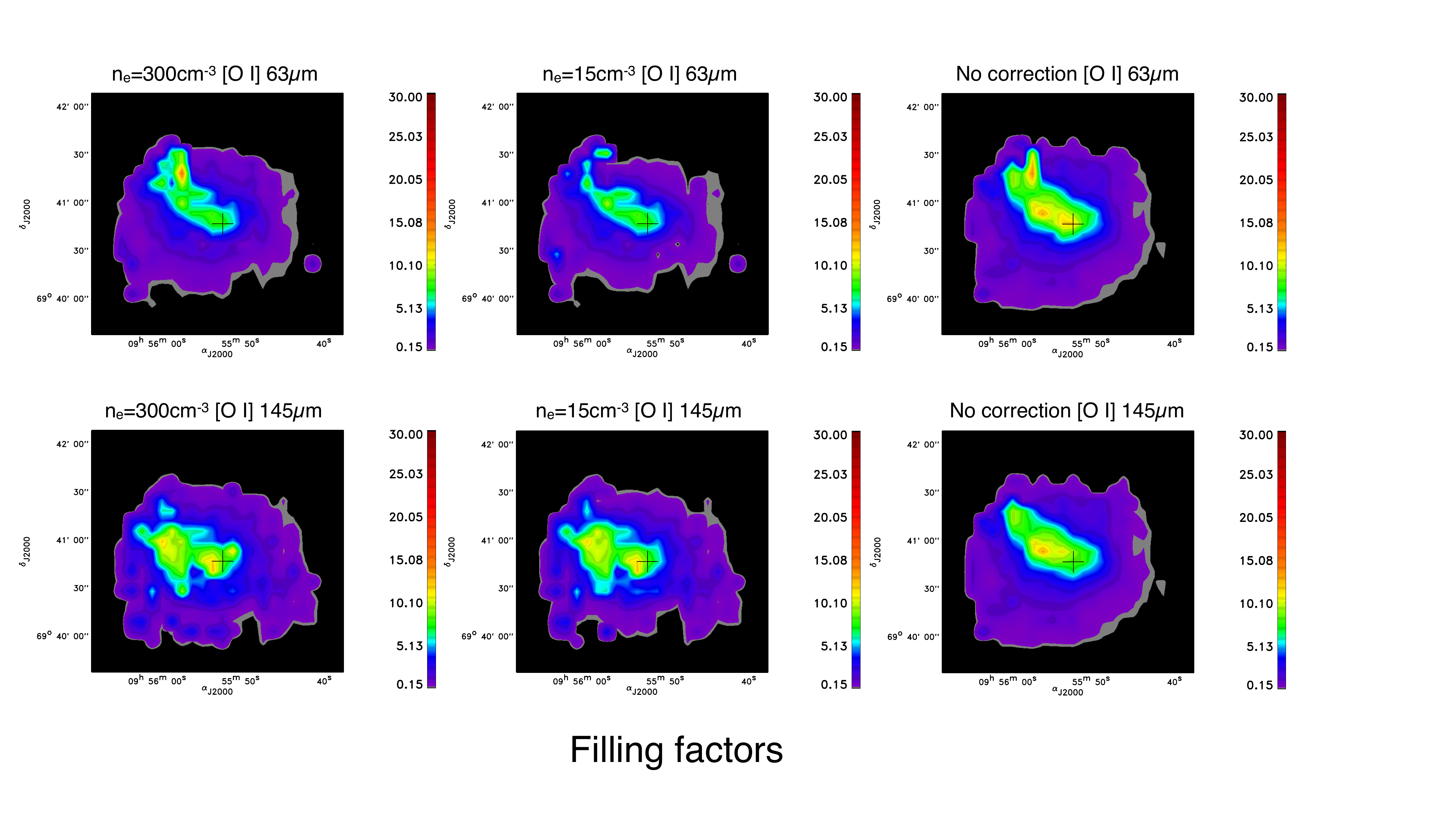}
       \caption{The  PDR solution maps obtained for  the area beam filling factor   $\phi$,   {\it i.e} the
       percentage of the beam covered by the emitting clouds. }
         \label{PDRsol_phi}
   \end{figure*}

Figure  \ref{Gnplot}  shows the logarithmic plot of $G_0$ versus $n_H$ for all solutions obtained with the input sets which included the [OI]
line at 145 $\mu$m.    Similar plots derived using the results  obtained with
the re-aligned [OI] 63
$\mu$m map, are consistent with those shown in Figure  \ref{Gnplot}. 
The macro regions defined for the different corrections to the [CII] total flux
are sketched in the top right panel. The same color code is used in the three plots shown in the figure. 
The upper panel shows the results
obtained using as input to the model the observed [CII] emission without any correction;
the bottom left (right) panel shows  the results obtained assuming that the electron 
 density of the ionized gas producing [CII] to be subtracted to the observed flux is 
is equal to 300 (15)  $\rm{cm^{-3}}$. 
The similarity between all  relationships  is evident, reinforcing our assertion that the correction 
to [CII] does not significantly impact our results.
This figure shows that $n_H$ and $G_0$  are correlated and that the correlation agrees with
what  was previously found by Malhotra et al. (\cite{Malhotra}) by modeling  a sample of local star-forming 
galaxies observed with the Infrared Space Observatory (ISO-LWS) and that are shown as yellow asterisks in Figure \ref{Gnplot}.
Also shown in this Figure is   the value obtained from Colbert et al. (\cite{Colbert}) from ISO--LWS  observations of M82. 
We find that the solutions for M82's starburst region corresponds well to the PDR solution determined 
by Malhotra et al.  and  Colbert et al. from the ISO data.
The ISO-LWS observations are single pointings centered on the galaxy centers 
(i.e do not encompass the entirety of the disk) and therefore are biassed towards the warm and dense 
central regions. This explains why these measurements occupy the high $G_0$--$n_H$ part of the diagram.
{\it {The results of this work extend the correlation by almost 2 orders of magnitude to much lower $G_0$--$n_H$}}.
All four components of M82 that we considered, follow the same correlation in $G_0-n_H$ as normal galaxies {\it suggesting a common origin
of the neutral gas emission for all regions, including the outflow}. This means that the neutral gas as traced by its main
coolants is consistent with arising by classical PDRs also in the outflow}.\\

\subsection{Column density and [CII] optical depth of the atomic neutral medium associated with the PDRs}
The averaged cloud column density of the hydrogen nuclei   associated with the [CII] 
emission, $N_{[CII]}(H)$, can be calculated using  the following equation (Crawford et al., \cite{Crawford}):

\begin{equation}
N_{[CII]}(H)  =\frac{4.25\times 10^{20}}{\chi(C)}
\left[\frac{1+2\times e^{(-92/T)}+(n_{crit}/n_{H})}{2~ e^{(-92/T)}}\right]~{\left[\frac{I_{[CII]}^{PDR}}{\phi}\right]},
\end{equation}

where $N_{[CII]}(H)$ is in cm$^{-2}$, $I_{[CII]}^{PDR}$ in erg
s$^{-1}$ cm$^{-2}$~ sr$^{-1}$, $\phi$ is the area filling factor, {\it i.e.} the  fraction of the beam filled 
with clouds emitting [CII], $\chi(C)$ is the [C$^+$]/[H] gas--phase   abundance ratio
that is equal  to 3$\times$10$^{-4}$ (Crawford et al \cite{Crawford}), $n_{crit}$ is the critical density for collision of C$^+$
with H atoms equal  to 4$\times$ 10$^3$ cm$^{-3}$,  T is the gas temperature in K and
n$_H$ is the gas density.  
Using the observed (and corrected) [CII] emission, the temperature, the  gas density and filling factor values resulting from the PDR modeling  
{\it {we are able  for the first time  to derive  the averaged cloud column density map of the HI gas associated with the PDRs  
emitting the observed [CII] 
in M82}}.\\
The resulting column density ranges from $\sim 1\times 10^{21}~\rm{cm^{-2}} $ up to $\sim 1\times 10^{22}~
\rm{cm^{-2}}$ for all input data
sets  except for  the ones obtained using  the [OI] line at 145 $\mu$m and [CII] 
not corrected for the ionized gas contribution, for which the  column density reaches  few 
$10^{22}~\rm{cm^{-2}}$.
We also  derived the [CII] optical depth map  by applying the following formula
(Crawford et al. \cite{Crawford}):
 
\begin{eqnarray}
 \tau_{[CII]} &=&  \frac{\lambda^3~A_{ul}}{8\pi\Delta v}
 \left[ \left(1+\frac{n_{crit}}{n_{H}} \right) e^{92/T} - 1\right] \times   \nonumber \\
             &  &   \left(\frac{2~e^{(-92/T)}}{1+2~e^{(-92/T)}+(n_{crit}/n_{H})}\right)N_{[CII]}(H)
\end{eqnarray}

The results obtained  are shown in Figure \ref{tau_CII}. 
 The morphology of both the average cloud column density (not shown here) and the [CII] opacity maps, 
 are obviously similar. All maps   have [CII] optical depths much less than 1. 
 The last column of Table \ref{Table1} lists the  average  $\tau_{[CII]}$  values of the starburst, the  disk , 
 the north  and south part of the outflow.   
 The map obtained with no correction to the observed  [CII] line flux
 shows  an opacity  higher (up to $\sim$ 0.1) than those obtained correcting the [CII] observed flux from ionized gas contribution,
  as expected, while the morphology is very similar in all three cases.
 Figure \ref{tau_CII} shows that the opacity is higher along the disk of the galaxy with a decrease toward the starburst
   region, where the brightest H$\alpha$ emission is also located. This opacity decrease could be the result
of   material cleared out from the winds powering the outflow. The opacity drop  is even more pronounced in the outflow  regions, especially
in the northern one.}  \\

Some estimates of the column density associated with the [CII] line emitting gas already exist for M82.
In particular we compare our results with those obtained from Crawford et al. (\cite{Crawford}). These authors
 derived $\tau_{[CII]}\lesssim 0.2$ and column  density  $N_{CII} \geqslant 9\times 
10^{17} ~ \rm{cm^{-2}}$ by using equation 3 and 4  in the limit of high density ($n \gg n_{crit}$) and
high temperature  ( $T \gg 92$ K). We have calculated a corresponding  column density and optical depth 
averaging the values
obtained from our  maps, in  a region of M82 which corresponds, as close as possible, to
the position and beam size used in the  Crawford observations.   We find an average [C$^+$] column density  a few 
$10^{18} ~\rm{cm^{-2}}$ and a opacity $\sim 0.03$,  in good agreement with the
limits   given by Crawford et al. (\cite{Crawford}) on these  parameters.

   \begin{figure*}
   \centering

\includegraphics[angle=0,width=18cm,height=15cm]{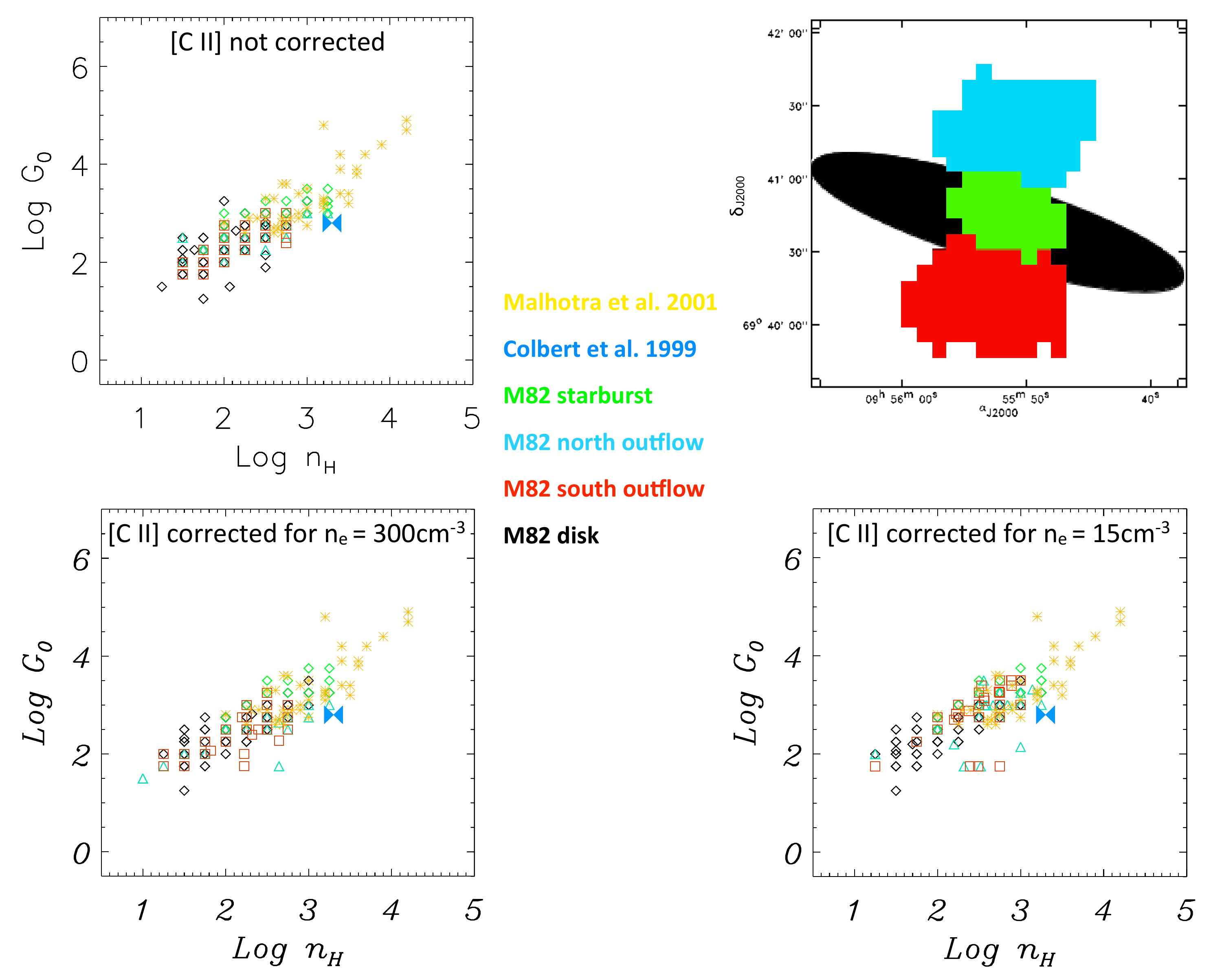}
        \caption{PDR solutions  for M82 obtained by using   [CI],[OI] at 145 $\mu$m and TIR as input parameters to the model. On the
	right of the top panel a schematic view of the masks used to define  the starburst (green), the disk
	(black),
	and the  north (cyan) and south (red) outflow is shown. The same color code has been used for the plotting.
	Also overplotted for comparison,   the solutions obtained for a local sample of normal star forming galaxies from
        Malhotra et al. 2001 (yellow stars) and with  the ISO--LWS 
       single pointing flux of M82   (Colbert et al.  1999). Top:  PDR solutions obtained
       by using as input the [CII] emission not  corrected for the ionized gas contribution. 
      Bottom left :   the results obtained by using the [CII] corrected for the ionized  gas contribution assuming a electron density equal to
      300 $\rm{cm^{-3}}$. Bottom right: the results obtained by using the [CII] corrected for the ionized  gas contribution assuming 
      a electron density equal to
      15 $\rm{cm^{-3}}$. (See section 4.1.3 for details).}
         \label{Gnplot}
   \end{figure*}

\subsection{Outflow   energetics  }
We can estimate the minimum mass of the hydrogenic nuclei associated with the
[CII] emitting gas using equation 1 of Hailey-Dunsheath et al. (\cite{Steve10}),

\begin{equation}
\frac{M_{[CII]}(H)}{M_{\odot}} =0.77\times \left(\frac{0.7 L_{[CII]}}{L_{\odot}}\right) \times \left(\frac{1+2 e^{(-92/T)}+n_{crit}/n}{2 e^{(-92/T)}}\right),  
\end{equation}

In what follows, we calculate the  mass, the kinetic energy and the mass outflow rate associated with the neutral atomic gas emission 
obtained in this work (Table\ref{Table2}) and we   compare them to the corresponding values of the cold molecular gas traced by 
the CO emission, the warm molecular gas traced by the $H_2$ emission and the ionized gas emitting in $H\alpha$
(Table  \ref{Table3}).\\
   Using equation 5 we calculate the neutral atomic gas {\it {masses associated with the outflow}}  for each input data set involving the [OI] line
at 145 $\mu$m only  (listed in Column 1 of Table \ref{Table2}).   The overall uncertainties for these quantities 
are comparable with those quoted for the PDR parameters.   \\

The total atomic mass in form of PDRs in the outflow is 
   $~2-8\times10^7~M_{\odot}$ depending on the input data,   a factor   $\sim$ 4-15 times lower than the 
cold molecular mass entrained in the outflow  as measured by Walter, Wei\ss~ and Scoville ( \cite{Walter}, $M \sim 3.3 \times 10^8~ \rm{M_{\sun}}$) 
and 3 orders of magnitude higher than the warm molecular gas (Veilleux, Rupke and
Swaters \cite{H2}).  
Our values should be regarded as lower limits 
because the area on which we have calculated the   outflow masses, 
is not defined by the intensity of the emission but by the extent of the mapped area. 
In other words, there may be still material further out that we
have just not mapped. Moreover, this is not the total atomic mass carried in the outflow but only that associated with the [CII]
emitting gas.
Therefore the atomic and molecular gas in the outflow could be considered as almost comparable. For comparison, the
mass of the warm filaments detected at optical wavelengths is $5.8 \times 10^6 ~\rm{M_{\odot}}$ (Shopbell and Bland--Hawthorn
\cite{Shopbell}).

The {\it {kinetic energy}} associated with the neutral gas   is  E$_{kin}$= M$_{[CII]}(H) \times v^2$ where
$v^2$ is the deprojected mean velocity of the outflow calculated as explained in section 3.3.   
We obtained values in the range of  $\sim 1-5 \times 10^{54}~\rm{erg}$ (Table \ref{Table2}, Column 2) 
for the two cones of the outflow.
From  Table \ref{Table3} we   see that these values are 1 order  of magnitude lower than what is found from the outflowing
molecular gas (Walter, Wei\ss~ and  Scoville   \cite{Walter})
and H$\alpha$ gas (Shopbell and Bland--Hawthorn \cite{Shopbell})   equal  to
$\sim$3$\times$10$^{55}$  erg. These values are also  $\sim$ 2 orders of magnitude higher than
the kinetic energy of the warm H$_2$ molecular gas calculated  
 by assuming that the velocity of the warm molecular gas is equal to 
that of the CO molecular gas (Veilleux , Rupke and Swaters \cite{H2}). Therefore, although not energetically dominant, the atomic gas 
in the outflow  still represents a significant contribution to its overall energetic.

We define  the {\it {dynamical time}}  of the outflow  as the time  necessary for 
its material  to travel  at  a mean velocity equal to $v_{deproj}$ up to  a distance of
450 pc away from the disk along the minor axis. 
We use this length because this is the extent
of the mapped area  above and below  the galaxy's disk.
We can apply this simple calculation to both the neutral and the ionized gas, the latter being traced by the [OIII] line
emission. The dynamical  time for both cones of the outflow is  $\sim$ $5\times 10^{6}$ yr     for the
neutral and the ionized  gas,
basically because they all have comparable deprojected mean velocities in the outflow.
This time is also comparable  to that estimated for the warm molecular H$_2$
by Veilleux , Rupke and Swaters (\cite{H2}) for the same distance we have assumed in our calculation. 

Finally we can calculate the  {\it {mass outflow rate}} of the neutral gas in the outflow  dividing the masses 
by the dynamical times.   For each input data set, the mass outflow rates  (Column 3 of Table \ref{Table2}) are similar in the
northern and southern cones of outflow. 
The range of the total   neutral gas  mass rate is 10-25 $\rm{M_{\sun}/yr}$.
However, if we add the mass outflow rate of the molecular component in the outflow (33 $\rm{M_\odot/yr}$, Walter, Wei\ss~ and Scoville
\cite{Walter}) the total PDR outflowing  mass rate  in the  outflow is comparable  to   
 the  total Star Formation Rate (SFR) of M82 ($\sim$ 25 $\rm{M_{\sun}/yr}$ ) estimated by
 F\"orster--Schreiber et al. (\cite{Natascha03}) giving a mass load ( defined as $\rm{\dot{M}_{Outflow}/SFR}) \sim 2$   
 similar to  that   measured in another starburst galaxy NGC253 (Sturm et al. \cite{Sturm11}).

   \begin{figure*}
   \centering
\includegraphics[angle=0,width=18cm,height=4.5cm]{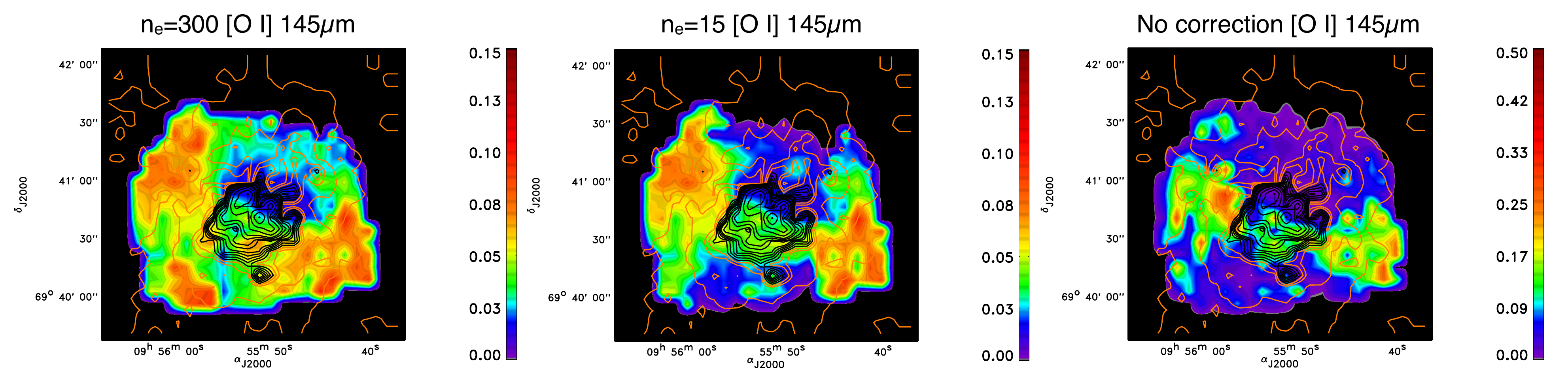}
 
       \caption{  The  [CII] opacity maps  obtained by using equation 4. Only the result of the modeling obtained using as input 
       data  set [CII] (with and without corrections), FIR and  the [OI] line at 145 $\mu$m are shown.}
         \label{tau_CII}
   \end{figure*}
 
\begin{table*}
\caption{Atomic gas characteristics in M82 outflow}              
\label{Table2}      
\centering                                      
\begin{tabular}{c c c c c c}          
\hline   
input set                                                  & $M_{[CII]}(HI)$    & $E_{kin}$     & $\dot{M}$        & \\
                                                           & $10^7 ~\rm{M_{\odot}}$  & $10^{54}~\rm{erg}$ & $\rm{M_{\odot}/yr}$   & \\
\hline   

\hline   
[OI] 145 $\mu$m, [CII]   corrected n$_e$= 15 $\rm{cm^{-3}}$ &                    &                &                 & \\
  North                                                     &     2.1            &     1.5        &      4.0        &     \\   
  South                                                     &     2.9            &     1.7        &      5.1        &     \\   
\hline   
[OI] 145 $\mu$m, [CII]   corrected n$_e$= 300 $\rm{cm^{-3}}$ &                   &                &                 & \\
  North                                                     &     4.3            &     3.2        &      8.4        &     \\   
  South                                                     &     8.4            &     5.0        &     14.8        &     \\   
\hline   
[OI] 145 $\mu$m,[CII] NOT corrected                         &                    &                &                 & \\
  North                                                     &     4.1            &     3.1        &      8.1        &     \\   
  South                                                     &     6.1            &     3.6       &      10.7        &     \\   
\hline                                             
\end{tabular}
\end{table*}

\begin{table*}
\caption{Masses, kinetic energies and mass outflow rates in the outflow of M82
for the cold, warm and ionized gas traced by the H$\alpha$ emission, obtained in other works. These values should be
  compared with the values obtained for the neutral atomic emitting gas calculated in this work  and listed in Table 1.}              
\label{Table3}      
\centering                                      
\begin{tabular}{c c c c c c c}          
\hline   
Gas phases                          & $ M_{[CII]}(HI)$      &    $E_{kin}$                      & $\dot{M}$        &   Reference \\
                                    & $ \rm{M_{\odot}}$          &      erg                         & $\rm{M_{\odot}/yr}$   &             \\
\hline   
Cold Molecular gas                  &	  3.3$\times 10^8$  &	  3.3$\times 10^{55}$	        &     33  	   &Walter, Wei\ss~ and Scoville 2002	        \\   
\hline   
Warm Molecular gas                  &	  1.2$\times 10^4$  &	  $10^{51}$ 	                &     0.001 	   & Veilleux, Rupke and Swaters 2009	      \\   
\hline   
Ionized gas traced by H$\alpha$     &	  5.8$\times 10^6$  &	  2.0$\times 10^{55}$ 	        &     3.6 	   &Shopbell and Bland--Hawthorn  1998	      \\   
\hline   
Neutral atomic gas                  &	 2-8$\times 10^7$  &	  1-5$\times 10^{54}$           &    10-25	  &	This work      \\   

\hline                                             
\end{tabular}
\end{table*}

   \begin{figure*}
   \centering
\includegraphics[angle=0,width=17cm,height=6.5cm]{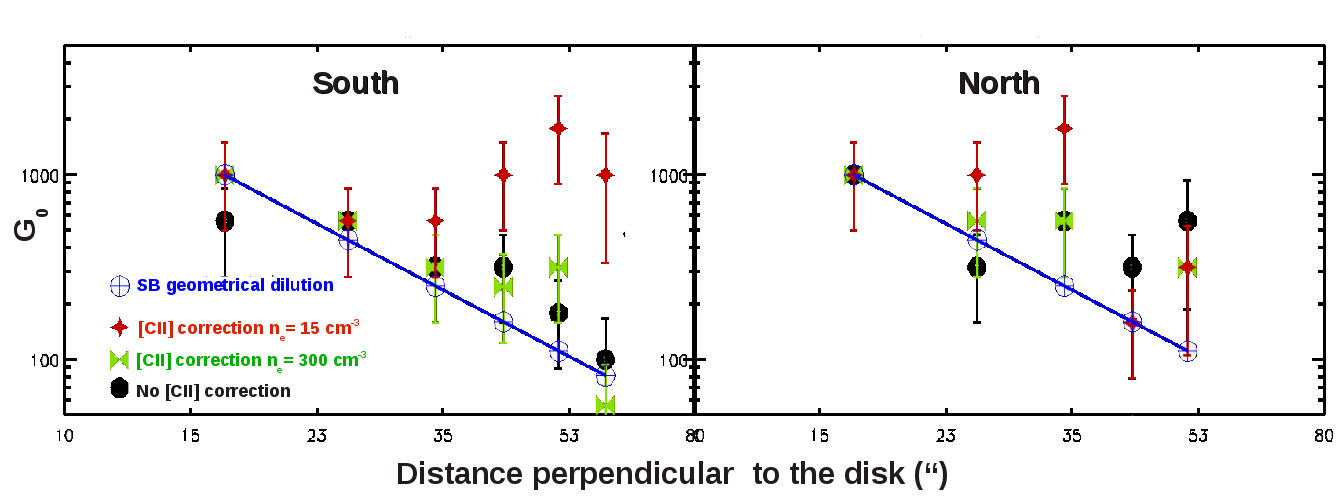}
 
       \caption{ The observed FUV ISRF in Habig units along the minor axis of M82 are plotted as a function of the distance from the galaxy
       center in $\arcsec$, for the south (left) and north (right) outflow separately. The blue symbols connected by the blue line, 
       are the values  calculated in the pure geometrical dilution hypothesis. The starburst region, which occupies the central 250 $\rm{pc}$
       ($\sim$ 15 $\arcsec$)
       is not included. }
         \label{Outflow_cut}
   \end{figure*}

\subsubsection{Comparison amongst the various phases of outflowing  material}
We can summarize the ISM phases participating to the  M82's outflow  as follows:
the  hot ($10^7$ K)  gas emitting in X-rays;  the ionized cooler gas ($10^4$ K) emitting in $H\alpha$;
 the cold and warm molecular gas (Walter, Wei\ss~ and Scoville \cite{Walter}, Veilleux , Rupke, and Swates \cite{H2}),   
 the cold  ( $\sim 300-400$ K) neutral gas (this work) 
  and the dust (Alton, Davies and Bianchi \cite{Alton}, Leeuw and  Robson \cite{Leeuw}, Kaneda et al. \cite{Kaneda}, Roussel et al. \cite{Roussel}). 
  These components are not all dynamically coupled to each other.
The  cold   molecular gas traced by the CO emission 
and the neutral gas traced by the  FIR fine structure line emission studied
in this work   have  similar averaged velocities and opening angles which suggests they are coupled.
 These components are  decoupled from the
ionized gas emitting in $H\alpha$ which has much higher velocity (Greve \cite{Greve}). 
 If we assume that  also the 
  PAHs emission observed by Engelbracht et al. (\cite{PAH}) and the warm molecular gas observed by 
  Veilleux , Rupke and Swaters (\cite{H2}), for which we do not have kinematic information,
  are also coupled with the [CII] emitting gas in the outflow,
  we have gathered  all fundamental components of the classical PDRs. \\
  Probably, the cold gas material 
  is entrained from the  disk into the outflow by the fast hot wind, in the form of  clouds. Once in the outflow, 
   the  UV photons in the wind
  itself and/or directly from the starburst region, excite the   external shells to these clouds which emit  in the 
  observed FIR fine structure lines, the aromatic bands from PAHs and   warm molecular gas
  traced by $H_2$ in the near infrared. 
  Micelotta, Jones and Tielens (\cite{Micelotta}) have shown that PAHs 
  can survive the  destruction by collisions with thermal electrons  in the  hot winds only if they 
  are in form of cold  cloudlets. This supports the idea of having the molecular and cold neutral medium 
  in  the outflow   organized in small clouds dragged
  by the wind, dynamically decoupled   from    the ionized gas traced by $H\alpha$ emission, but almost equally energetically 
  important. \\
    
The analysis of  the behavior of the derived $G_0$ along the minor axis of M82 can help test  
  this   hypothesis.   
  In Figure \ref{Outflow_cut} we show the values of $G_0$ we have obtained from the PDR modeling along the minor axis of
the galaxy as a function of the distance from the galaxy's center, for the different input data sets we have used. 
The two panels show the cut for the southern  (left) and   the northern (right) outflow.\\
We compare these values to the ISRF values provided by the starburst assuming  pure geometrical dilution without any extinction.
This corresponds to the maximum possible contribution of the starburst light in the outflow.
In order to calculate it we assume the following:
 1) the starburst has a diameter of 500 $\rm{pc}$ (F\"{o}rster Schreiber et al.  \cite{Natascha03}), which corresponds
to $\sim$30 $\arcsec$, and therefore we consider only the points at distances $\gtrsim $15$\arcsec$, along the minor axis on each side; 
2) the ISRF at   this distance in both cones of the outflow, is equal to what
we have  calculated from the PDR modeling at the same distance.  The resulting values are indicated with the blue  
solid line in both panels of figure \ref{Outflow_cut}.
This figure  shows that the ISRF values derived from the PDR modeling agree  within the uncertainties  with the blue 
line,  except for the $G_0$ values obtained by using as input to the PDR models the [CII] corrected from   the ionized gas contribution 
for a density equal to 15 $\rm{cm^{-3}}$, which is a rather extreme case. Agreement with the maximum Go obtained by neglecting extinction would indicate that  
most of the starburst FUV photons can  travel   the outflow, with a low probability of being absorbed. 
Since we know that there is dust in the outflow, 
the only way this can happen is that there are big dust-free holes or, in other words, 
that most of the existing   dust   must be concentrated in 
clouds with  low filling factors. This supports  the cloudlets interpretation we have proposed  above  and 
it also shows  that the starburst light is
sufficient to heat most of these cloudlets  and that no significant extra heating (i.e. {\it in situ} star formation) is required.\\

  We have already discussed in Section 3.6   that the ionized gas traced by the [OIII] line at 88 $\mu$m and that traced
  by the H$\alpha$ emission are not kinematically coupled and probably not even co-spatial in the outflow. 
  We have also shown evidences that these two lines likely trace  gas ionized by two different physical processes: the H$\alpha$
  emission arises from a combination of
photoionization by the starburst and shocks: the first process dominates the emission  towards the outflow axis, the second process 
  traces predominantly the gas shocked when the  hot flowing plasma encounters  the halo, and it is therefore more confined in the
    outflow walls. The [OIII] 88 $\mu$m emission is more concentrated along the minor axis of the galaxy and it 
    is most likely dominated by photoionization.
 Taking also into account that  the [OIII] 88 $\mu$m emitting gas flows at the same velocity of the neutral atomic gas, we propose
 that this is   gas   surrounding the PDR-molecular cloudlets ionized by the starburst light. This
  ionized component though is more collimated than the neutral gas component, suggesting that the ionized gas associated
  with the PDR clouds,  is decreasing towards the periphery of the outflowing material.     This is probably due to the fact that
    the number of UV photons necessary to excite the [OIII] line ($> 35.12~ eV $) directly emitted
from the starburst decreases faster than the number of the less energetic UV photons necessary to excite the [OI] and [CII],
 as the length of the material they have to pass through increases.\\ 
   Finally, there is yet a third component of material entrained in the
  outflow: that of the   dust detected through  the polarization of the  visible light reported by  
   Yoshida, Kawabata and  Ohyama (\cite{Yoshida}). This dust, which scatters the nuclear visible
   light, is made by grains larger and hence heavier than PAHs. In fact their velocity is lower 
   ($\sim$ 10 $\rm{km~s^{-1}}$ at 1 kpc distance above the galaxy's disk) than that observed for the PDR
   component associated with the cold gas and presumably with PAHs.

\subsubsection{ What   powers  the outflow in M82?  } 
In this Section we try to establish whether  the radiation pressure  or/and the mechanical energy due to the supernovae explosion
is powering the PDR component participating to the motion in  the outflow of M82 and whether the  outflow is  energy  or momentum 
driven ({\it i.e.} whether they conserve the energy or the momentum}).

The   {\it momentum rate provided by supernovae explosion} can be calculated following equation 10
 from  Murray, Quataert and Thompson (\cite{Murray}):

\begin{equation}
\dot{P}_{SN} \sim 2 \times 10^{33} \left(\frac{ SFR}{1~M_{\sun}~yr^{-1}}\right) ~~~ \rm{g~cm~s^{-2}}
\end{equation}

This assumes a supernovae rate $\nu_{SN}= 0.01~ \rm{yr^{-1}}$ for a SFR  equal to 
$1~\rm{M_{\sun}~yr^{-1}}$. F\"orster--Schreiber et al. (\cite{Natascha03})
have calculated that the supernovae rate in M82 is  0.13 $\rm{yr^{-1}}$ and that the SFRs of the two starburst episodes at
their
peak are equal to 18.5 and 6.3 $\rm{M_{\sun}~yr^{-1}}$  
giving a total SFR of  24.8 $\rm{M_{\sun}~yr^{-1}}$. 
A supernovae rate of $  0.01~ \rm{yr^{-1}}$ for a SFR equal to $1~\rm{M_{\sun}~yr^{-1}}$ corresponds to a supernovae rate of $\sim 0.25$
per a SFR of  24.8 $\rm{M_{\sun}~yr^{-1}}$, which is a factor of 2 higher than what found in M82.
This means that  we have to divide  equation 6  by 2. 
The total momentum rate available from the SN explosion  is then $\sim 2.5\times 10^{34}~  \rm{g~cm~s^{-2}}$.\\
The {\it  momentum rate given by the radiation pressure} is $L_{SB}/c$, assuming that the optical depth at the base of the outflow 
is 1.
The total bolometric starburst luminosity of M82 is $6.6\times 10^{10} ~\rm{L_{\sun}}$
(F\"orster--Schreiber et al. \cite{Natascha03}), which gives
a radiation pressure momentum rate equal to $8.5\times 10^{33}~  \rm{g~cm~s^{-2}}$, a factor of 3 less than
mechanical  momentum rate.\\
The {\it observed momentum rate} for both cones of the outflow ($\dot{M}_{\sun} \times v_{deproj}~ \rm{ g~cm~s^{-2}}$)  ranges from  5 to 12 $\times 10^{33} ~
\rm{g~cm~s^{-2}}$, depending on the input data set  (see Table \ref{Table2}). \\

The {\it energy rate due to supernovae explosion} under the same assumptions of equation  6 is
(equation 34 of  Murray, Quataert and Thompson \cite{Murray}):

\begin{equation}
\dot{E}_{SN} \sim 3\times 10^{40}*\left(\frac {SFR}{1~M_{\sun}~yr^{-1}}\right) ~~~ \rm{g~cm~s^{-2}}
\end{equation}

which for M82 translates in an energy rate $\sim 3.7 \times 10^{41}~ \rm{erg~s ^{-1}}$.\\
  The {\it observed energy rate} carried by the atomic PDR component in the outflow of M82 ($\frac{1}{2} \dot{M} v_{deproj}^2$) ranges from 2 -5
  $\times 10^{40}~ \rm{erg~s^{-1}}$.\\

 The momentum and energy  rates in the outflow carried by  the cold molecular component are equal to $21\times
 10^{33}~\rm{g~ cm~s^{-2}}$ and    $ 10^{41}~ \rm{erg ~s^{-1}}$  respectively,    
assuming a deprojected molecular velocity of 100 $\rm{km ~s^{-1}}$ and a dynamical
time equal to $10^7~yr$ (Walter, Wei\ss~ and Scoville  \cite{Walter}). 
We have seen in the previous section that  the  cold molecular component traced by the CO emission is likely to
be dynamically coupled to the neutral atomic gas  PDR component. Under this assumption we can add the
energetics of these two components to estimate a total balance between the momentum (energy)
 available and those observed to move this material in   the outflow.
We obtained a total observed momentum rate $\dot{P}_{atomic}+\dot{P}_{molecular} \sim 26-33\times 10^{33}~ 
\rm{g~cm~s^{-2}}$
and an observed energy deposition rate equal to $\dot{E}_{atomic}+\dot{E}_{molecular} \sim 12-15\times 10^{40}~
\rm{erg~s^{-1}}$ to be compared
with the    total momentum   and  mechanical energy deposition rates available of $\sim 34\times 10^{33}~ 
\rm{g~cm~s^{-2}}$
 and  $\sim  37\times 10^{40}~ \rm{erg~s^{-1}}$ respectively.\\
Table \ref{Table4} summarizes these  values: on the left it lists the momentum (energy) rate available in the system and
on the right the  observed momentum (energy) rates in M82.
Both the total momentum and  energy deposition rates  available in the system are comparable
to the observed values. The momentum driven outflow is not compatible with the observations if the source of power 
is the radiation pressure alone. 
  The energy driven case seems to leave  more room for agreement  between the observed and the available
  energy of the PDR outflowing material. \\
However, if we include in the energy and momentum budget calculation also the numbers  necessary to sustain the outflowing 
material observed in  $H\alpha$ by  Shopbell  and Bland-Hawthorn   (\cite{Shopbell}) the energy driven 
case becomes  only marginally compatible with the observations.
In fact, assuming a mass of the outflow observed at these   wavelengths  equal to $5.8\times 10^6~ \rm{M_{\sun}}$, a kinetic energy
equal to  $2.0\times 10^{55}$  erg, a mean deprojected velocity of ~$\sim 600 ~\rm{km~s^{-1}}$ ( Shopbell  and Bland-Hawthorn  
 \cite{Shopbell}) and a length of the outflow above the galactic plane equal to 1 kpc, one derives a
momentum rate $\sim 2.2\times 10^{33}  \rm{~g~cm~ s^{-2}} $ and an energy rate $\sim  40\times 10^{40} ~\rm{erg
~s^{-1}}$ (also listed in
Table \ref{Table4}). 
Taking into account the overall uncertainties ($\sim$ a factor of 2)  of the numbers derived from the above calculation 
and considering the cold  molecular gas, the neutral atomic gas and hot ionized gas phases energetic together,
we conclude that,  {\it assuming  a simple jet-like outflow geometry, both the energy and momentum driven process are compatible with the observed values, although the 
momentum driven case seems to be  slightly favored.}

\begin{table*}
\caption{Available and observed molecular and atomic gas energetics in the outflow of M82}              
\label{Table4}      
\centering                                      
\begin{tabular}{c c c c c c c c c c c c }          
\hline 
\hline 
&  Momentum Rates               &     Available                         &                 &            Used                    &\\ 
&                               &      $10^{33}~\rm{g~cm~s^{-2}}$            &                 &      $10^{33}~\rm{g~cm~s^{-2}}$        &\\

\hline 
& $\dot{P}_{SN}$            &     25                     &      $\dot{P}_{molecular}$	   &        21          &\\
& $\dot{P}_{SB}$            &     8.5                    &      $\dot{P}_{atomic}$         &       5-12         &\\
&                           &                            &      $\dot{P}_{H\alpha}$        &         2          &\\
\hline   
& {\bf{total}}               &    {\bf{33.5}}             &                                 &        \bf{28-35}          &\\
\hline 
\hline                                              
&  Energy Rates             &    Available            &                            &     Used   	        &\\ 
&                           &    $10^{40}~\rm{g~s^{-1}}$   &                            &      $10^{40}~\rm{g~s^{-1}}$	&\\ 
\hline                                              
&   $\dot{E}_{SN}  $	    &	    37                &     $\dot{E}_{moeluclar} $ &	  10	                &\\
&                          &                         &     $\dot{E}_{atomic}$     &	  2-5                    &\\
&                          &                         &     $\dot{E}_{H\alpha}$    &	  40                    &\\
\hline   
&   {\bf{total}} 	     &	   {\bf{ 37}}  	       & 			   &	   \bf{52-55}			&\\
\hline

\hline 
\end{tabular}
\end{table*}

\subsection{Line  diagnostic diagrams}
\begin{figure*}
   \centering
\includegraphics[angle=0,width=15cm,height=17cm]{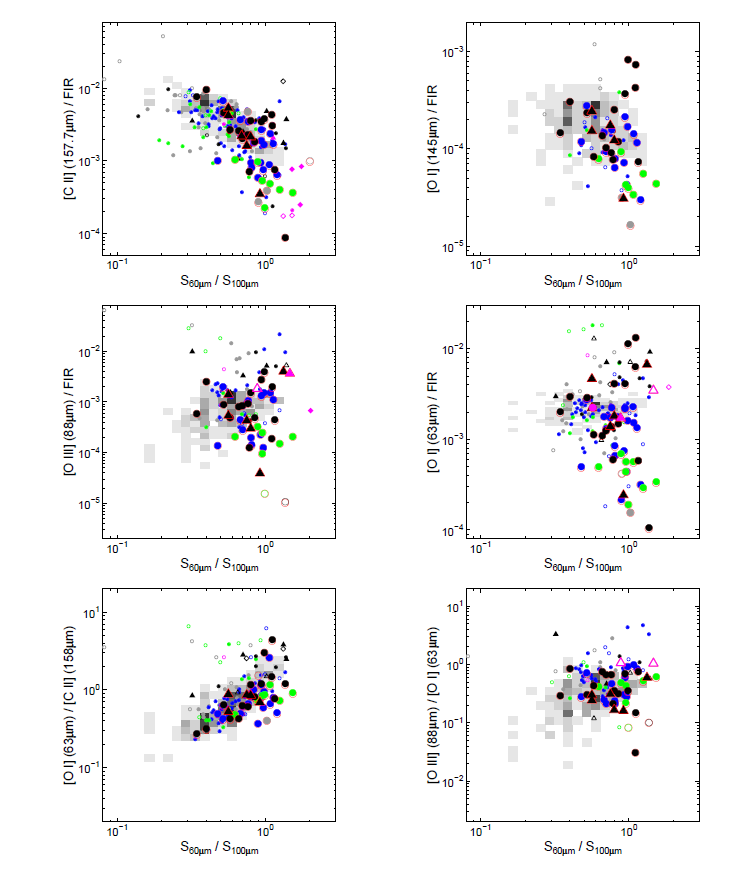}
       \caption{[CII]/FIR, [OI] 145$\mu$m/FIR, [OIII]/FIR,  [OI] 63 $\mu$m/FIR,  [OI] 63 $\mu$m/[CII] and [OIII] 88/[OI] 63 $\mu$m
        versus the continuum 60/100 $\mu$m ratio. M82 values are shown
       as gray scale. Symbols are the following (Graci\'a Carpio et al. \cite{Javier11} and in
       preparation):
       Big symbols: SHINING/PACS observations;
 Filled symbols: the lines are detected;
 Open symbols: one or both lines are not detected;
 Blue points: HII galaxies;
 Green points: LINER galaxies;
 Black symbols: AGNs;
  Circles: Seyfert 2 galaxies;
  Triangles: Seyfert 1 galaxies;
  Diamonds: QSOs;
 Magenta points: high-z galaxies;
 Grey points: unclassified galaxies;
 Small magenta points: flux at 25 or  at 100 $\mu$m is an upper limit. }
         \label{lineratioJavier}
   \end{figure*}

\begin{figure*}
   \centering
\includegraphics[angle=0,bb=1.0in 1.0in 7.5in 10in,width=18cm,height=23cm]{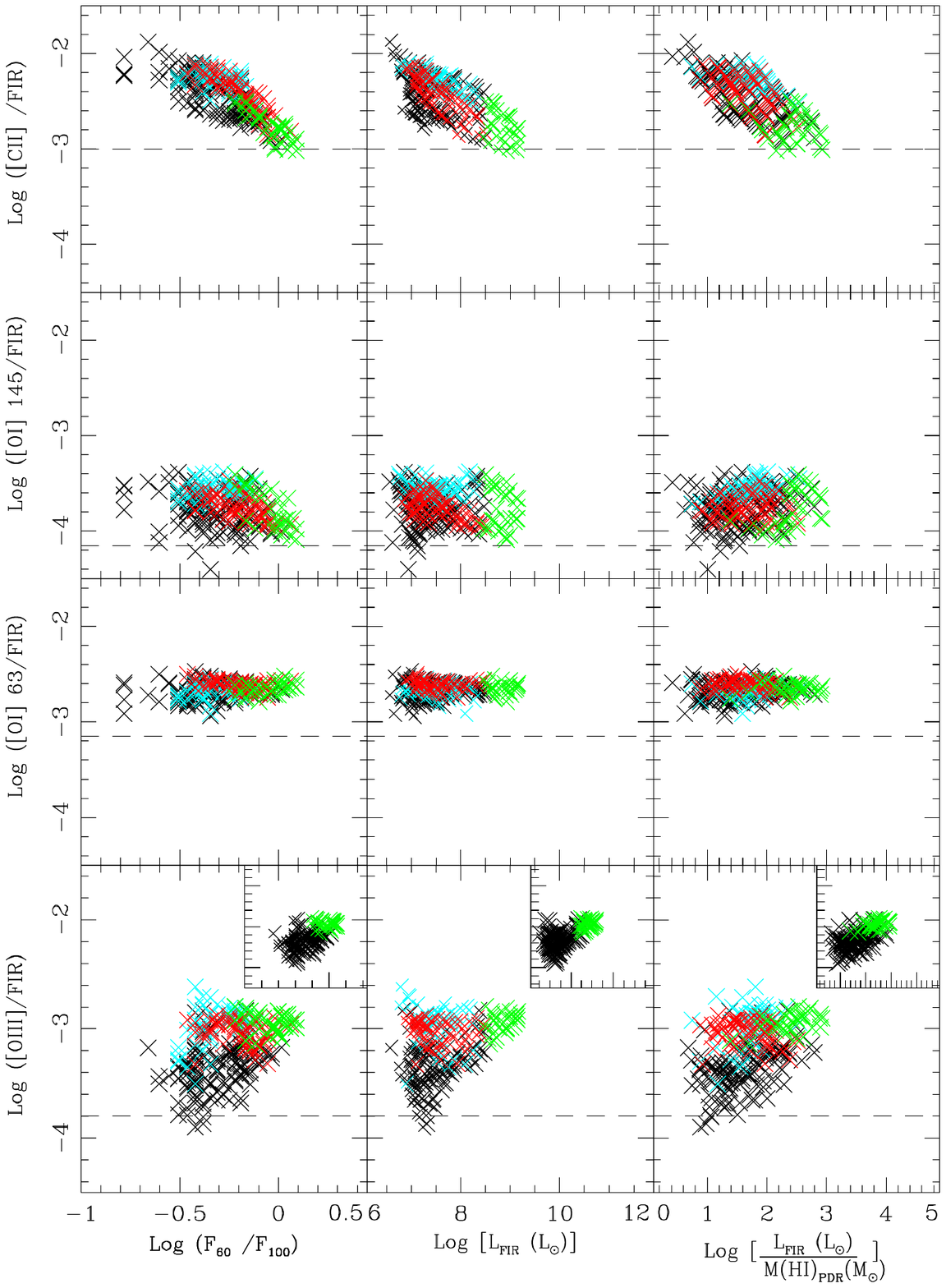}
       \caption{ The  Line/FIR ratio  versus: left: 60/100 colors; middle: $L_{FIR}$; right : $L_{FIR}/ M(HI)_{PDRs}$.
        Data are for M82 only. Black points are from the diffuse disk;
        red points from the southern outflow; cyan points from the northern outflow and green points from the starburst.
	The small  frames in the last row show the same relation reported in the big frames but only for the points
	belonging to the disk (black) and the starburst (green).
	  The   horizontal lines correspond  to the threshold identified by 
	Graci\'a Carpio et al.
(2011) between galaxy with normal (above) and 
deficient (below) $FIR_{line}/FIR_{continuum}$ ratios, or between normal and high efficency of SF.}    
         \label{lineratioJavier2a}
   \end{figure*}

Figure \ref{lineratioJavier} shows the ratio of each of the observed lines to the FIR continuum as well as the 
[OI] 63 $\mu$m/[CII], and [OIII]/[OI] 63 $\mu$m ratios as a
function of the $F_{[60\mu m]}/F_{[100 \mu m]}$ ratio, equivalent to the IRAS colors, derived from our PACS maps.
The latter  traces the dust temperature and therefore,  to  first order  
also the interstellar radiation field.
The M82 values are shown in gray scale. Overplotted are the data collected from
the literature or measured in the PACS GT Key Program SHINING (Graci\'a Carpio et al.
\cite{Javier11}). Different
symbols refer  to different galaxy types (see figure caption for details).
The emission line data points  of M82 are those derived from the final maps reduced 
to the same spatial resolutions. The
maps at  60 and 100 $\mu$m have been derived as explained in Section 4.1.1.\\
There is  excellent agreement between the M82 data, which probe regions of linear extent 
of $\sim 300$  pc, and the global  emission properties   of
galaxies. This means that   the physical  mechanisms governing the
observed relationships  act up to a scale  as small as 300 pc, and they are the
same independent on whether they act in  the starburst region, the diffuse disk
or the outflow. The only difference is in the panel
showing the [OI] 63 $\mu$m /FIR ratio that in the case of M82 is remarkably constant, while the galaxy
sample is exhibiting a large scatter. It is interesting to note that the  scale  we probe ($\sim  300~pc$)  is
  comparable to the  upper limit of the scale where the Kennicutt-Smidth law begins to break  (Schruba et al. \cite{Schruba}).\\
Each  left and middle panel of  Figure \ref{lineratioJavier2a}   shows  
the $FIR_{line}/FIR_{continuum}$ ratios versus the $F_{[60\mu m]}/F_{[100 \mu m]}$ colors  and the FIR luminosities  for    M82 only, color coded
depending on whether the regions belong to the outflow, the starburst or the disk of M82 as shown in the upper panel
of Figure \ref{Gnplot}.
The top  panels   show the [CII]/FIR ratio decreases by one order of magnitude with the increase of the $F_{[60\mu m]}/F_{[100 \mu m]}$ IRAS color
and the FIR luminosity, 
nicely following the decreasing trend outlined from the whole SHINING galaxy sample (Graci\'a Carpio et al. \cite{Javier11}). As expected, the lowest [CII]/FIR values in M82 arise from the
starburst region (green points).
This decrease was first discovered with the ISO LWS and it has been referred to as the 
 {\it {[CII]--deficit}}   
(Fischer et al.\cite{Fischer99}, \cite{Fischer2001},  Malohtra et al. \cite{Malhotra}, Luhman et al.
\cite{Luhman98}, \cite{Luhman}, Abel et al. \cite{Abel2009}), {\it i.e.} the warmer, more active and more infrared luminous   the systems are, the less [CII]
with respect to their FIR emission they emit. 
 
 The near universality of the [CII] deficit for luminous systems  has been  recently 
 challenged by
  observations of high redshift galaxies, which show a "normal" [CII]/FIR ratio despite of their 
   high FIR luminosities  (Maiolino et al. \cite{Maiolino}, Hailey-Dunsheath et al. \cite{Steve10},
    Ivison et al. \cite{Ivison},  Stacey et al. \cite{Stacey} ,Wagg et al. \cite{Wagg}, Sturm et al. \cite{Sturm2010}).
On the other hand,  if  the [CII]/FIR ratio  is plotted against the ratio between the 
FIR luminosity and the $H_2$ molecular mass, rather than FIR luminosities, the scatter in the relation 
is significantly reduced as is the disparity between systems at low and high redshift. 
This means that the [CII] deficit becomes universal at all redshifts above a threshold 
of the $L_{FIR}$/$M_{H_2}$ ratio, a commonly used indicator of star formation efficiency or evolutionary stage 
(Gonz\'alez-Alfonso et al, \cite{Eduardo}, Graci\'a Carpio et al.,  \cite{Javier11}).
Furthermore,  Graci\'a Carpio et al.  find that this deficit extends  to other PDR and pure HII region FIR lines (see their figure 2).
In galaxies where $L_{FIR}$ is predominantly powered by star formation, 
the  $L_{FIR}/M_{H_2}$ ratio is, to first approximation,
 the ratio between the energy released by the star formation  and the gas reservoir 
from where the stars form and therefore it is a parameter   directly related to the
star formation efficiency. This  is not the case for  
the $L_{FIR}$.   Graci\'a Carpio et
al. (\cite{Javier11}) show that above a certain $L_{FIR}/M_{H_2}$  threshold 
one also finds  an increase
 of the ionization parameter $U$, {\it i.e.} the ratio between the ionizing photons to the gas density at the inner surface of the ionized cloud. 
 The higher  this parameter is the larger  the HII region  and  the FIR emitted by the
 dust heated in the HII region. 
This increasing FIR does not mean a corresponding increase 
in the FIR emission lines arising from the PDR, thus the global $FIR(PDR)_{line}/FIR(HII+PDR)_{continuum}$ ratio decreases. 
 The values of the $L_{FIR}/M_{H_2}$ threshold ($>$ 80 $L_{\odot}~M_\odot^{-1}$ ) 
    is   very close to the threshold above which the star formation becomes much more efficient
  than what is predicted    by the   Kennicutt-Schmidt relation for normal star forming  galaxies  (Genzel
  et al.  \cite{Genzel10}).  This similarity suggests that the  galaxies and the regions within galaxies which 
  exhibit a  $FIR_{line}$ deficit, are in this highly efficient mode
of star formation. \\

In order to determine in which mode of  star formation the various regions of M82
are, we want to plot the line/FIR ratios of the points belonging to M82 as a function of   $L_{FIR}/M_{H_2}$.
Since we do not have a CO map of M82 available from which we can derive the $H_2$ mass,  we 
 use instead  the mass of the PDR gas  to calculate the  $L_{FIR}/M_{[CII]}(H)$  ratio. 
The cold molecular gas is more closely related to the site of star formation than the atomic neutral  gas, but
the atomic gas we are considering here is that associated with the PDR and therefore it is also related to the star
formation, although not as directly as the molecular gas. 
This is why the ratio between the FIR luminosity and the mass of neutral hydrogen 
associated with the PDRs can be considered  to first order,  as  star formation efficiency. \\
We have produced  plots analogous  to the $FIR_{line}/FIR_{continuum}$ deficit diagrams shown 
in Graci\'a Carpio et al. (\cite{Javier11}) for M82 (color coded: 
disk in black,   starburst in green, north outflow in cyan and south outflow in red, see also the schematic
figure on top right of Figure  \ref{Gnplot}). These diagrams 
are shown in  the right panels of Figures   \ref{lineratioJavier2a}  for each line.
We have also plotted the horizontal lines that Graci\'a Carpio et al. (\cite{Javier11}) consider as the limit between galaxy
with normal and deficient $FIR_{line}/FIR_{continuum}$ ratios. 

 The first thing to notice is
 that   {\it even the starburst region of M82  is not in this high efficiency mode of star formation}, 
 since  all points are above the horizontal lines. 
  Nevertheless we observe different behaviors among the different lines.\\
   The [CII]/FIR ratio gently decreases in all cases, even when   plotted as function of the  $L_{FIR}/M_{[CII]}(H)$, whereas
 the similar plot in Graci\'a Carpio et al.,  (\cite{Javier11}) who used the mass of H$_2$, shows  an almost constant
  behavior up to 
($\sim$ 80 $L_{\odot}~M_\odot^{-1}$ ), and then a decrease. However, we notice  that the data shown in   Graci\'a Carpio (\cite{Javier11})
did not include the whole SHINING sample\footnote{The SHINING sample, P.I. E. Sturm,  includes 50 galaxies ranging from HII, starbursts, Seyfert1 and 
Seyfert2, and LINERs and including LIRGs and ULIRGs,  Graci\'a Carpio et al. (in preparation)} and the data available at that time contained more ULIRGs/AGNs than starbursts
and star forming   galaxies.  
Now that the sample is complete and  more starbursts/star forming galaxies  populate the low   $L_{FIR}/M_{H_2}$ part of the plot, 
a   decrease  
 at low   $L_{FIR}/M_{H_2}$ is also clearly visible  for the SHINING sample
 (Graci\'a Carpio et al. in preparation)
 although with a  slope  much less steep than  that followed by   galaxies at $L_{FIR}/M_{H_2} > 80$. 
 This continuous decrease is predicted by models which assume  an increasing production of FIR 
emission arising by the dust heated in  
  the HII regions, as the ionization parameter U increases (see Figure 12  of Abel et al.   \cite{Abel2009}).\\
The [OI] 63  $\mu$m /FIR and the  [OI] 145  $\mu$m /FIR  ratios in M82 are remarkably constant  and, in contrast to the [CII]/FIR behavior,  
do  not show any hint of decline in M82. The constancy is compatible with  what is observed in the whole SHINING sample, which shows a decline 
of the [OI]/FIR ratio only at 60/100 ratios higher than that probed in M82. \\

Both trends can be understood by modeling  these ratios with   CLOUDY as   shown in   
Figure 3 of  Graci\'a Carpio et
al. (\cite{Javier11}). These figures show  the   $FIR_{line}$/FIR ratios  on diagrams of   the ionization parameter $U$ versus   the gas  density $n_H$.
Each diagram is divided into gray and red zones, corresponding to the typical values of starburst/star forming galaxies and ULIRGs respectively.
The loci of the line over FIR value corresponding to the transition between the zones   correspond  to the horizontal lines shown in 
Figure \ref{lineratioJavier2a}.
First we address the diagrams involving only the PDR lines ( [CII], and the two [OI] lines): when moving diagonally on these diagram,
 {\i.e. increasing both parameters},   the [CII]/FIR ratio decreases but the [OI]/FIR ratios do not change 
significantly because in the first case one moves perpendicular to the loci of  constant [CII]/FIR and in the second case 
  one moves almost parallel to the loci of constant [OI]/FIR, at least in the range of line/FIR ratios spanned by M82 and starbursts in
general (gray zones in Figure 3 of  Graci\'a Carpio et
al.  \cite{Javier11}). 
 The behavior  of the lines arising from ionized gas is reversed with respect to what we have just examined in the case of PDR lines. 
 In particular  we concentrate here on the [OIII]/FIR line loci,
on the U-$n_H$ diagram. We can see that in this case, increasing  both parameters means increasing the HII-line/FIR ratio, as it is observed 
in M82 shown in the bottom panel of  figure \ref{lineratioJavier2a}. However, it is worth  mentioning that the pure HII region lines  are very sensitive to 
the input SED used for the modeling, and in particular to the hardness of the exciting photons, much more than the
lines originating from PDRs or the dust temperature and total flux which is sensitive  to the energy of the heating photons rather than to the hardness. \\
From the $U$-$n_H$ diagram  we can derive the $FIR_{line}/FIR_{continuum}$ ratios expected for the  values calculated 
by  F\"{o}rster Schreiber et al., \cite{Natascha01} for M82 ($U=-2.3$ dex and $n_e~ \sim n_H = 300 ~ \rm{cm^{-3}}$). These values
 agree well  to the averaged values we observe, as shown by the green crosses in all panels of \ref{lineratioJavier2a}.\\

  Finally, a closer look to the trend followed by the $[OIII]/FIR$ ratio in Figure \ref{lineratioJavier2a} seems to suggest 
that there exist   two different behaviors in M82. 
In the disk (black points) and in the starburst (green points) this ratio increases with IRAS color, FIR luminosity 
and  $L_{FIR}/M_{H_2}$ as expected,  (see inner panels in  the bottom row in Figure \ref{lineratioJavier2a}). 
On the other hand, in the outflow this ratio remains constant and  is
 higher than the corresponding point of the disk at low IRAS color, FIR emission and $L_{FIR}/M_{H_2}$ ratios.
 This suggests that in the outflow the [OIII] is enhanced with respect to the FIR emission. We have seen that the [OIII]
 emission in the outflow is dynamically coupled with the gas emitting in the [CII] and [OI] lines, that we have
 associated to PDR clouds. We  have also seen that the PDRs in the outflow survive in form of small cloudlets. We can
 imagine that the dust  associated with these small clouds is shielded from the UV photons resulting in 
 colder FIR colors  and lower FIR emission. We notice that shocks do not play a significant role in the heating 
 of the PDR material  anywhere in M82 including the outflow, since the [OI] 63 $\mu$m /[CII] ratio is always well below the values ($\sim
 10$) predicted in the presence of shocks (Hollenbach and McKee \cite{Hollenbach1989}).

\section{Summary and Conclusions}
We have mapped a 2.5$\arcmin$$\times$2.5 $\arcmin$ region of M82 in four FIR atomic fine-structure lines, 
[CII] at 158 $\mu$m, [OI] at 63 and 145 $\mu$m and [OIII] at 88 $\mu$m, 
with the PACS spectrometer on board the Herschel Space Observatory, reaching unprecedented 
spatial resolution ($\sim 200 -300$ pc) at these wavelengths. 
The mapped area covers approximately 1 kpc along the minor axis of the galaxy. 
This enables detection of the M82 outflow structures in these lines for the first time.  
Since the [CII] and [OI] lines are the main coolants of the cold neutral medium, these observations 
trace and probe the atomic gas in the outflow.  This paper presents the first analysis of these 
observations with emphasis on the nature and structure of the atomic gas in the outflow and on its relation with 
other ISM phases studied in previous works. We summarize the main results as follows.
 
 \begin{itemize}
\item{We derived intensity, radial velocity and dispersion maps of the entire mapped 
region in all lines.  We also used the offline continuum images and the parallel array 
scans to produce a dust temperature map.  The line dispersion maps most clearly show two 
dynamically separated structures, disk and outflow, since the dispersion in the outflow 
is $\sim 50- 100 ~\rm{km~s^{-1}}$ higher than that in the disk. These distinct components are also visible 
in the radial velocity maps and indicate that the northern outflow is receding from us while the 
southern outflow is moving toward us, in agreement with previous results.  
In the outflow, the mean  deprojected velocity, calculated  assuming a simply inclined jet-like geometry,
is  $\sim75~ \rm{km~s^{-1}}$  for both  the cold neutral   and the ionized gas traced by the [OIII] 88 $\mu$m line, 
 close to the average   velocity of the molecular gas ($\sim100~ \rm{km~s^{-1}}$)
 reported by Walter, Wei\ss~ and Scoville (\cite{Walter}) and much smaller than the velocity of 
 the outflowing material observed in $H\alpha$ ($\sim 600 ~\rm{km~s^{-1}}$,   Shopbell  and Bland-Hawthorn  
 \cite{Shopbell}) }

\item{Analysis of the line ratio maps and of the line-to- continuum maps confirms earlier 
findings that the base of the northern outflow is more obscured than the southern one due 
to the inclination of the galaxy disk. These maps also reveal that the opening angle of the 
cold gas component of the outflow is much larger than the ionized gas traced by the [OIII] 88 $\mu$m 
which appears much more 
collimated. } 
   \item{ We have carried out PDR modeling of the neutral gas emission lines to derive 
   the key physical parameters of the atomic gas within and surrounding the starburst, 
   stellar disk and the outflow components of M82. By applying the Kaufman et al. (\cite{Kaufman}) 
   PDR models to each pixel of our line intensity images we were able to produce, for the first time 
   in an external galaxy, maps of the FUV interstellar radiation field $G_0$, the gas density $n_H$, 
   the gas temperature $T_{gas}$, the atomic gas cloud beam filling factor, the column density of atomic hydrogen,
    and the [CII] opacity. Since the correction for the contribution of [CII] emission from ionized gas is 
    very uncertain we run the model with and without it, but the resulting maps do not differ significantly. 
     Solution maps using each [OI] line together with the [CII] and continuum TIR emission were 
     similar to within the errors for all parameters except the gas density,   probably because of alignment
     problems between the [OI] 63 $\mu$m and [CII] map. For this reason and also because the [OI] 63 $\mu$m 
     optical depth   
     can vary significantly   in the galaxy we based our analysis only on the results obtained
     using as input to the model the [OI] line at 145 $\mu$m together with [CII] and TIR. The area beam  filling factor decreases  
 by up to   an order of magnitude in the diffuse disk and in the outflow. $G_0$ and $n_H$ both vary  from
 $\sim 50$ to few $10^3$   and the atomic gas temperatures are in the range of $300 - 400~ ^\circ$ K.
   The [CII] optical depth is everywhere much less than unity, enhanced along the disk but decreasing toward the starburst region,
 perhaps because some material has been cleared out by the  winds that power  the  outflow.}
\item{  
 The relation between $G_0$ and $n_H$ in M82 is  the same as has been observed in previous works for the global emission of
 star forming galaxies    extended   by almost two orders of magnitude
   towards smaller values. In this plot, the same relation is followed regardless of whether the points belong
 to the starburst, the outflow or the diffuse disk,  confirming that the clouds in the
 outflow are still organized in PDRs very similar to the clouds in the rest of the galaxy.}
\item{  
  We find that  the mass and the kinetic energy of the outflowing cold neutral gas are 
 $\gtrsim 2-8\times 10^7~ \rm{M_{\sun}}$  and  $\gtrsim 1-5 \times 10^{54}$ ~erg    slightly
 smaller than  the 
 corresponding values found for the outflowing molecular gas ($\sim 3.3\times 10^8~ \rm{M_{\sun}}$,
 $\sim 3 \times 10^{55}$  erg). The mass outflow  rate of the molecular and atomic gas is 
$\sim$ 43-58 $\rm{M_\odot yr^{-1}}$ comparable to the SFR of M82 equal to 25 $\rm{M_\odot yr^{-1}}$,    resulting 
in a mass loading factor  ${\dot{M} }$/SFR $\sim$ 2.  
 This result,  together with the fact that these  two components  show similar velocities, suggests that they are
 dynamically coupled with similar origin. Since the cold atomic   gas is consistent with
 emission from classical PDRs ({\it i.e.} interface regions between the ionized and molecular gas) with small filling
 factors,
the coupling of these two cold components of the ISM is naturally explained if both the molecular and atomic
 media belong  to the same disk clouds  entrained in the outflow by the wind where they 
partially evaporate due to heating driven by thermal conduction from the hot gas, 
surviving as clouds  smaller   than their original size.   This scenario is supported by the fact that 
  the decline of the derived FUV ISRF  in the outflow  is consistent with a pure geometrical dilution, 
  indicating that FUV photons can travel in the outflow with a low probability of being absorbed by dust, 
  which therefore must be concentrated in small clouds. This also implies  that in the outflow, 
  there is  no need for a  significant {\it in situ} star
  formation in addition to the starbust light.}
 
 \item{We have found that in the outflow, the ionized gas traced by the FIR [OIII] 88 $\mu$m line flows at a much smaller 
 velocity ($\sim  75 ~\rm{km~s^{-1}}$) than the ionized gas traced by the H$\alpha$ emission 
 ($\sim 600~ \rm{km~s^{-1}}$, Shopbell and Bland--Hawthorn  \cite{Shopbell}) but at a
 velocity similar to that of the neutral atomic gas traced by the [CII] and [OI] lines. We have also not found   
 line splitting in any of the observed FIR lines, including the [OIII] line at 88 $\mu$m, despite the fact that 
 the line splitting detected in H$\alpha$  is large enough to be detected at the PACS spectral resolution. 
 These results
 suggest that [OIII] 88 $\mu$m and H$\alpha$ line emission  trace two different ionized gas components,   both in
   velocity and in   physical  space. We   propose  a scenario in which the bulk of the [OIII] line at 88 $\mu$m
 traces   gas surrounding the PDR cloudlets observed  in the [CII] and [OI] lines, photoionized by the starburst radiation. 
 This would justify the fact that all
 these components flow  at the same velocity. In this scenario, the bulk of the H$\alpha$ 
 emission arises from the shocked gas at the interface between the hot plasma wind emitting in X-ray and the halo.
  This would explain why the H$\alpha$ emission is mainly located in the walls of the outflow as demonstrated by the observed
  line splitting.}

\item{The momentum and the energy available to drive the outflowing gas, molecular  atomic and ionized,
are both compatible with that observed in the flows, although   the momentum driven case seems to be slightly 
favored by the data. }

\item{We compare the trends found in the relative FIR fine-structure line fluxes as a 
function of the IRAS 60/100 colors of normal and infrared-bright galaxies to those found in the spatially resolved, 
$\sim 300$ pc regions of M82.  We find that for all but the warmest far-infrared colors of ULIRGs and 
the most extreme LIRGs, the ratios in M82 overlap with the trends found recently by 
Graci\'a Carpio et al. (\cite{Javier11}) for the whole galaxies.  These authors explore parameter space in density 
 and ionization parameter.  Based on photoionization models, they argue that the atomic fine-structure line 
 deficits found in the neutral gas of galaxies with the warmest FIR colors and the highest values 
 of  $L_{FIR}/M_{H_2}$ are caused by the presence of high ionization parameters in which the dust rather than the gas 
 absorbs most of the ionizing photons.  They suggest that these most extreme galaxies are in the high 
 efficiency mode of star formation rather than in a normal/Milky Way-like mode of star formation.  
 Importantly we find that down to 300 pc size regions, even in regions of the highest star formation 
 efficiency in M82, only the normal mode of star formation is present, i.e. no hint of bimodality is found.
}

\end{itemize}

\begin{acknowledgements}
PACS has been developed by a consortium of institutes led
by MPE (Germany) and including UVIE (Austria); KU
Leuven, CSL, IMEC (Belgium); CEA, LAM (France); MPIA
(Germany); INAF-IFSI/OAA/OAP/OAT, LENS, SISSA (Italy);
IAC (Spain). This development has been supported by the
funding agencies BMVIT (Austria), ESA-PRODEX (Belgium),
CEA/CNES (France), DLR (Germany), ASI/INAF (Italy), and
CICYT/MCYT(Spain). \\
Basic research in IR astronomy at NRL
is funded by the US ONR; J.F. also acknowledges support from
the NHSC.\\
E.G-A thanks the support by the Spanish Ministerio de Ciencia e Innovaci\'on 
under project AYA2010-21697-C05-01, and is a Research Associate at the 
Harvard-Smithsonian Center for Astrophysics.

\end{acknowledgements}

\begin{appendix} 
\section{}
\begin{figure*}
   \centering
\includegraphics[angle=0,width=19cm,height=11.0cm]{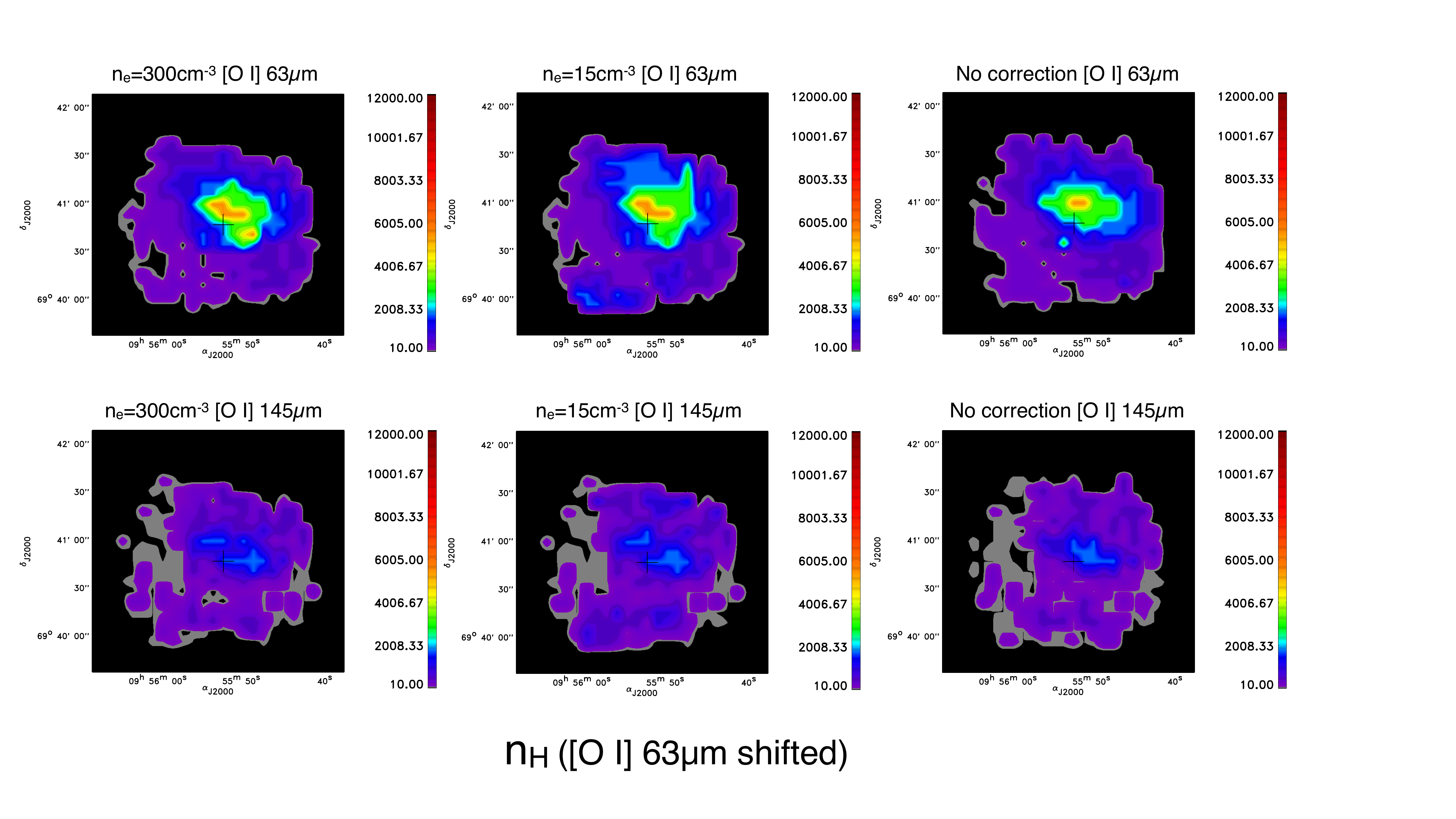}
       \caption{The  PDR solution maps obtained for the gas density $n_H$ using the [OI] map at 63 $\mu$ shifted to match the steep gradient
       of the starburst region to that in the [CII] map, instead of the [CII] peak emission. }
         \label{PDRsol_nH2}
   \end{figure*}

We have seen in Section 4.2 that we find a significant difference in the morphology of the gas density map 
obtained using as input to the PDR
modeling  the datasets with the two [OI] lines (Figure \ref{PDRsol_nH}), and this difference is independent
 on the correction 
we apply to the observed [CII] emission due to the   contribution from ionized gas.
If this difference were caused by the fact that [OI] at 63 $\mu$m is  thick in these regions, 
 we would expect  to observe less [OI] 63 than what  is predicted by the solution obtained using  the [OI] line at 145  $\mu$m, which is
 optically thin. We have checked this and we find the opposite answer, that is  in order to make the two density map  solutions similar,
  we should {\it reduce}  the  input flux at 63 $\mu$m. In other words, it looks like, in these regions we have too much 
[OI] 63 $\mu$m   flux.  Although this could be related to the uncertainties of the geometrical factor equal to two we have applied 
 to the observed [OI] 63 line  flux (see section 4.1.2), we wanted to check whether this difference could  arise  from our 
 processing steps. \\
 Each set of input 
 values to the PDR modeling were obtained from   maps (either [CII], FIR and [OI] at 63 $\mu$m or [CII], FIR and [OI] at 145 $\mu$m),
 rebinned to the largest pixel size (6$\arcsec$, that in the red channel), smoothed to the worse resolution (that of the [CII] map), 
 reduced to the same size and aligned to the [CII]  peak.
 In this chain of processing, the map which suffers   the heaviest manipulation is that at 63 $\mu$m. This is because the original pixel
 size is smaller, and its PSF differs from that at 158 $\mu$m  much more than that at 145 $\mu$m.
 Moreover, the observations at 158 and 145 $\mu$m were carried out in the same AOR, therefore the  pointing accuracy  is likely to affect
 these two observations in a similar way. This may give rise to alignment problems between the final map at 63 $\mu$m
 and those at higher wavelengths.\\ 
 For these reasons we decided to run the models again, this time  using the [OI]  63 $\mu$m  map aligned   on the [CII] map 
 on the steep gradient 
 of the flux distribution rather than to the peak (shift of 1 pixel = 6$\arcsec$).  Figure \ref{PDRsol_nH2} 
 clearly shows that now both gas density maps show an  enhancement toward the north part of the starburst, although the values are lower for
 the solutions obtained with the [OI] 145 $\mu$m line than those obtained with the [OI] 63 $\mu$m line. The maps in the other parameters,
  $G_0$, $T_{gas}$ and $\phi$, show the same morphology as those shown in figures \ref{PDRsol_G0}, \ref{PDRsol_Tgas} and
  \ref{PDRsol_phi}. 

\end{appendix}

\end{document}